\documentclass[11pt]{article}
\usepackage{geometry}                
\usepackage{natbib}
\usepackage{graphicx}
\usepackage{amsmath,amssymb,mathabx}
\usepackage{epstopdf}
\usepackage{lineno}
\usepackage{authblk}
\usepackage{setspace} 
\usepackage{appendix}
\usepackage{color}
\usepackage{setspace}
\usepackage{appendix}
\def\jgr{{J. Geophys. Res.}}
\def\gji{{Geophys. J. Int.}}

\def\jfm{{J. Fluid Mech.}}
\def\pepi{{Phys. Earth Planet. Inter.}}

\def\eq#1{\begin{equation}#1\end{equation}}
\def\eqlbl#1{\label{eq:#1}}
\def\eqref#1{(\ref{eq:#1})}
\def\dd{\mathrm d}
\def\v#1{{\bf #1}}
\def\Di{\mathcal D}
\def\Dt#1{{{\rm D}#1\over {\rm D}t}}
\def\nablab{{\mbox{\boldmath $\nabla$}}}

\title{Remarks on compressible convection in Super-Earths}
\author{Yanick Ricard}
\author{Thierry Alboussi\`ere}
\affil {Universit\'e de Lyon, ENSL, UCBL, UJM, Laboratoire LGLTPE,
15 parvis Ren\'e Descartes, BP7000, 69342, Lyon, Cedex 07, France.}

\begin{document}
\maketitle
\begin{abstract}
The radial density of planets increases with depth due to compressibility, leading to impacts on their convective dynamics. To account for these effects, including the presence of a quasi-adiabatic temperature profile and entropy sources due to dissipation, the compressibility is expressed through a dissipation number, $\Di$, proportional to the planet's radius and gravity. In Earth's mantle, compressibility effects are moderate, but in large rocky or liquid exoplanets (Super-Earths), the dissipation number can become very large. This paper explores the properties of compressible convection when the dissipation number is significant. We start by selecting a simple Murnaghan equation of state that embodies the fundamental properties of condensed matter at planetary conditions.
Next, we analyze the characteristics of adiabatic profiles and demonstrate that the ratio between the bottom and top adiabatic temperatures is relatively small and probably less than 2. We examine the marginal stability of compressible mantles and reveal that they can undergo convection with either positive or negative superadiabatic Rayleigh numbers. Lastly, we delve into simulations of convection performed using the exact equations of mechanics, neglecting inertia (infinite Prandtl number case), and examine their consequences for Super-Earths dynamics.
\end{abstract}

\section{Adiabatic conditions inside a convective planet}

It is well known that convection of a compressible fluid at high Rayleigh number brings the average radial profiles of density,
temperature and pressure close to their adiabatic and hydrostatic values ($\rho_a$, $Ta$, $P_a$) according to
\begin{subequations}
\begin{align}
{\dd \ln{\rho_a}\over \dd z} +{\alpha_a g\over  \Gamma_a C^a_{P}} =0,\eqlbl{adiaq1} \\
{\dd \ln{ T_a}\over \dd z} +{\alpha_a g\over  C^a_{P}} =0,\eqlbl{adiaq2}\\
{\dd P_a\over \dd z} +\rho_a g  =0, \eqlbl{adiaq3} 
\end{align}
\end{subequations}
$z$ being the vertical coordinate (directed against gravity $\v{g}=-g\v{e}_z$). In these equations, $\alpha$ is the thermal expansivity, $C_P$ is the heat
capacity (or specific heat) at constant pressure  and $\Gamma$ is the Gr\"uneisen parameter  
\eq{\Gamma={\alpha K_T \over \rho C_V}={1\over \rho C_{V}}\left({\partial P\over \partial T}\right)_V.\eqlbl{gruneisen}}
The superscript or underscript 'a' in equations \eqref{adiaq1}-\eqref{adiaq2}-\eqref{adiaq3} indicates that the various quantities are computed
along the adiabat itself.

A reasonable equation of state (EoS) for a condensed planet is based on the observation that 
the Gr\"uneisen parameter
is essentially a function of density \citep{anderson79} according to
\eq{\Gamma=\Gamma_0\left({\rho_0\over \rho}\right)^q,\eqlbl{grurho}}
where $q$ is around 1, $\rho_0$ and $\Gamma_0$ are the density and Gr\"uneisen parameter at standard conditions that we choose to be at the surface of the planet. On average, the Gr\"uneisen parameter 
is between 1 and 2 in the mantle \citep[e.g.][]{stacey04} or in the core \citep{alfe}. 
In the following we will use $q=1$. In this case the EoS appropriate for condensed materials becomes
\eq{P={K^0_T\over n}\left[\left({\rho\over\rho_0}\right)^n-1\right]+\alpha_0 K^0_T (T-T_0),\eqlbl{EoS1}}
where at reference temperature $T_0$, the pressure-density relation is given by a Murnaghan expression \citep{murnaghan51} and $n\approx 3-4$ for solid silicates and for liquid silicates or metals.
In the equation \eqref{EoS1}, $\alpha_0$ and $K_T^0$ are the thermal expansivity and the isothermal incompressibility at reference conditions.
Although EoS \eqref{EoS1} is simple and empirical, it encapsulates the typical properties of solids and fluids
and gives a very good fit to the radial density of the Earth assuming its adiabaticity, away from the transition zone discontinuity \citep{Ricard22}. This EoS implies direct relations between thermal expansivity and incompressibility with density which are
\eq{\alpha=\alpha_0 \left( {\rho_0\over \rho}\right)^n, \eqlbl{alpha}}
\eq{K_T=K_T^0 \left( {\rho\over \rho_0}\right)^n, \eqlbl{K}}
and again these two expressions provide realistic expressions of the properties measured in laboratory
experiments.
Since the Gr\"uneisen parameter \eqref{grurho} is only density-dependent,  
one gets a simple relation between the adiabatic temperature and the adiabatic density by combination of \eqref{adiaq1} and \eqref{adiaq2}
\eq{T_a=T^t_a \exp \left[ {\Gamma_0} \left( {\rho_0\over \rho^t_a}-{\rho_0\over \rho_a} \right) \right], \eqlbl{Ta0}}
where $T^t_a$ and $\rho_a^t$ are the surface ($t$ stands for top) adiabatic temperature and density.

We make two additional approximations when deriving analytical expressions  for the adiabatic profiles (those approximations will not be used in the numerical computations as they would invalidate them). 
\begin{itemize}
\item We assume the heat capacities at constant volume and temperature $C_V$ and $C_P$ are equal and constant. The heat capacites
are indeed very close when $\alpha T <<1$ which is the case for condensed material, and they are both close to the Dulong and Petit
values   $C_V\approx C_P\approx 3\mathcal{R}$ in J K$^{-1}$ mol$^{-1}$ ($\mathcal{R}$ is the gas constant) at large temperatures \citep{dulongpetit}.  
\item The surface adiabatic 
temperature is likely different from the surface temperature, however as $\alpha (T_a^t-T_0)<<1$,  the surface adiabatic density and the reference density can also be identified, $\rho_a^0\approx \rho_0$ (in other words the density of planets is a function of pressure, not temperature).
\end{itemize}
At last assumption is made on the variation of gravity
with depth in a generic planet. For simplicity, we assume that gravity is uniform which is basically the case in Earth's mantle.

With these hypotheses it is easy to solve for the adiabatic conditions in a layer where $z$ varies between $0$ and $H$ (e.g., a mantle of thickness $H$, $z=0$
being at the core-mantle boundary), and we get
\begin{subequations}
\begin{align}
\rho_a   &=\rho_0  \left(1+{H-z\over h} \right)^{1/(n-1)}, \eqlbl{rhoadia} \\
P_a       &={n-1\over n}{ \rho_0 g h} \left[ \left({\rho_a\over \rho_0}\right)^{n} -1\right], \eqlbl{Pa} \\
T_a       &=T^t_a \exp \left[ {\Gamma_0} \left( 1-{\rho_0\over \rho_a} \right) \right] \eqlbl{Ta}
\end{align}
\end{subequations}
where
\eq{h= {1\over n-1} {K_T^0\over   \rho_0 g}={1\over n-1} {\Gamma_0  \over \Di} H. \eqlbl{h}}
In the last equality, we have introduced the dissipation number $\Di$ defined by
\eq{\Di={\alpha_0 g H\over C_V}. \eqlbl{D}}
This surface dissipation numebr is only expressed from quantities known at the surface, this choice seems to be the only possible choice when exploring a new planet.
In the Earth the dissipation number is around $\Di_\Earth =0.71$ in the mantle and 0.56 in the liquid core  (using $\alpha_0=3\times 10^{-5}$ K$^{-1}$, $H=2900$ km and $C_V=1200$ J K$^{-1}$ kg$^{-1}$ in the mantle,  $\alpha_0=1.8 \times 10^{-5}$ K$^{-1}$ \citep{murphy13} , $H=2300$ km and $C_V=715$ J K$^{-1}$ kg$^{-1}$  \citep{gubbins} in the liquid core, with $g=9.8$ m s$^{-2}$).

 In geophysical textbooks \citep[see, e.g.,][]{schubert} $\Di$ is defined with a $C_P$ in the denominator which does not make much
practical difference as their difference is always neglected in the geology literature.
However  we prefer to defined the dissipation with $C_V$
like in the definition of the Gr\"uneisen parameter. We are  free to assume that one of the two  heat capacities is a constant
(here $C_V$ is chosen constant), but assuming the constancy of both heat capacities leads to inconsistencies 
in the energy conservation \citep{albou13} because their difference is directly related to the EoS
through Mayer's relation.

The previous   equations \eqref{rhoadia}-\eqref{Pa}-\eqref{Ta} can be used to discuss the possible characteristics of the adiabatic profiles of large planets. From the variety
of masses and radii of exoplanets that have been detected it seems that many of them are rocky at least until a radius
of order 2.5 times the Earth's radius  \citep{otegi20}. Their observed mass $M$ increases
roughly as a power 3.45 of their radius $R$ (their large internal pressures increase their average densities as $\approx R^{0.45}$). We will use this observation to scale
the gravity in our equations with $g\propto  M/R^2 \approx R^{1.5}$ and we consider that the thickness of the convective layers is proportional to $R$.
With these scalings, the dissipation number $\Di$ varies like $gH\propto R^{2.5}$. Therefore, according to \cite{otegi20}
dissipation numbers up to $2.5^{2.5}=10$ times that of the Earth can be expected in fairly common rocky planets, and in the following we will explore
dissipation numbers up to $\Di=10$. We will use $\Di=\Di_\Earth (R/R_\Earth)^{2.5}$ when, to set ideas, we 
discuss in term of planetary radii instead of dissipation numbers;  a Super-Earth with a radius twice that of the Earth (resp. 3 times) 
would therefore be assumed to have a mantle with a dissipation number around 4.0 (resp. 11.1) and a core with a dissipation number around 3.2 (resp. 8.7).

\section{Adiabatic conditions in a Super-Earth}
\subsection{The adiabatic density and temperature profiles}
\begin{figure}
\centerline{\includegraphics[width=9cm,angle=0]{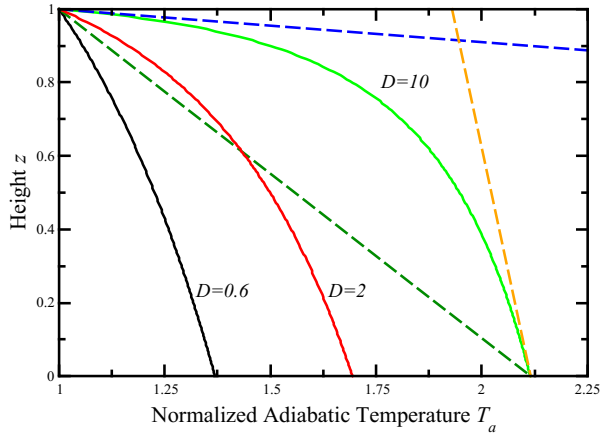}}
\caption{
Normalized adiabatic temperature for $\Di=0.6$, 2, 10 (black, red, green). In the case $\Di=10$, the dashed lines are conductive profiles that will be discussed below, one is with $\Delta T=\Delta T_a$ (dark green), one
with a gradient equal to the bottom adiabatic gradient (orange), the last one (blue) carries at the surface the same heat flow as the adiabatic gradient at the surface.
}
\label{figTa}
\end{figure}

As the incompressibility increases very significantly with density and therefore with pressure, the adiabatic density and temperature only increase moderately as a function of the planetary radius. 
In Figure \ref{figTa}, we depict the adiabatic temperature  normalized with its value at the surface according to \eqref{Ta}. We use $\Di=\Di_\Earth=0.6$ (black), $\Di=2$ (red), $\Di=10$ (green).

\begin{figure}
\centerline{\includegraphics[width=9cm,angle=0]{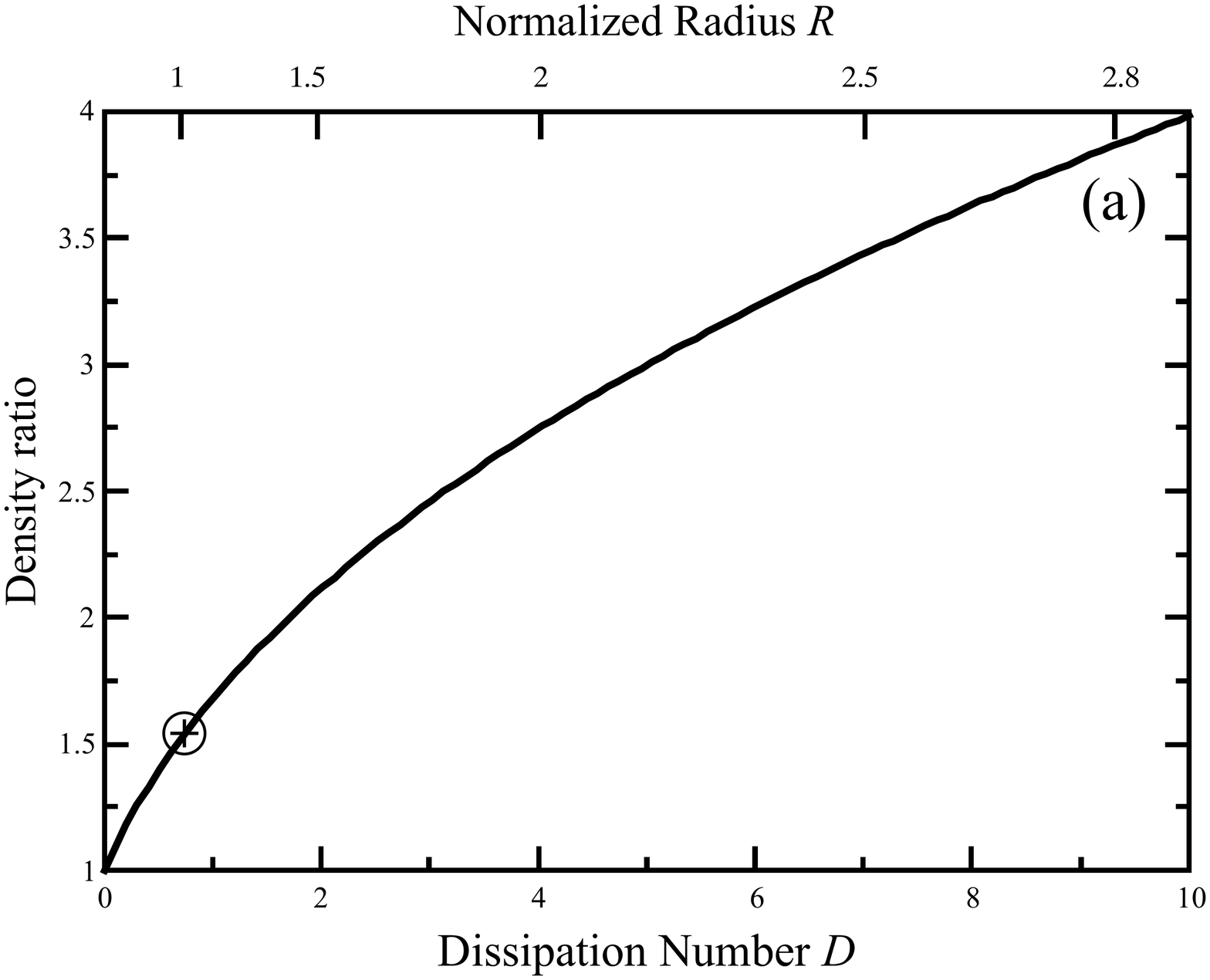}~\includegraphics[width=9cm,angle=0]{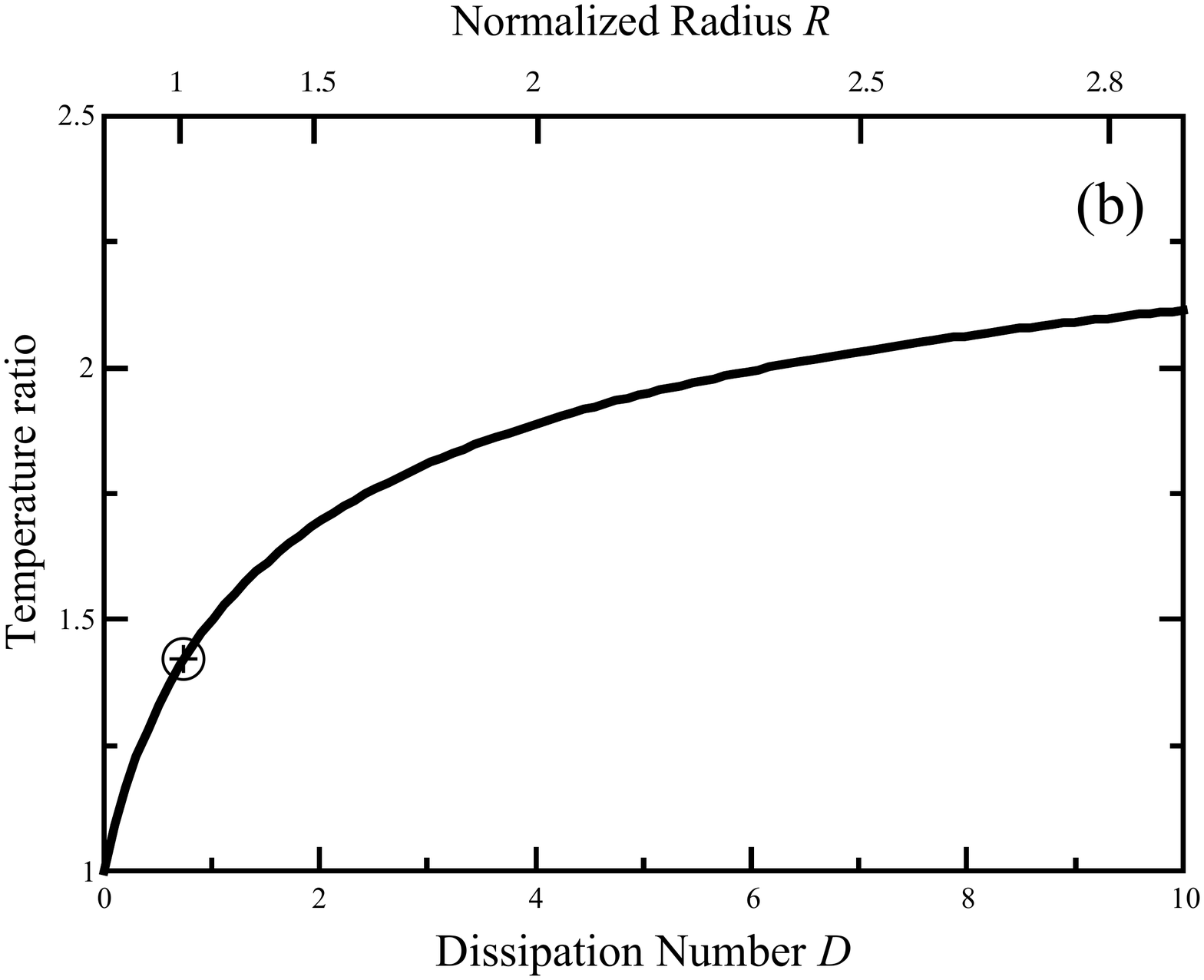}}
\caption{
Ratio between bottom and top adiabatic density (panel a) and temperature (panel b) in a compressible planet as a function of dissipation $\Di$ ($n=3.3$, $\Gamma_0=1$).
The symbol $\Earth$ indicates the situation for the Earth. The horizontal axis is either labelled in dissipation numbers (bottom) or in Super-Earth radii (top).}
\label{Ratio}
\end{figure}

The ratio of the adiabatic density between the bottom and the top of a convecting mantle is according to  \eqref{rhoadia} and \eqref{h}
\eq{{\rho_a^b\over \rho_a^t}  = \left(1+(n-1) {\Di\over \Gamma_0}\right)^{1/(n-1)}. \eqlbl{rhojump}}
This ratio in plotted in Figure \ref{Ratio}a as a function of $\Di$ (bottom axis) and as a function of $R$ (top axis, assuming $\Di\propto R^{2.5}$).
The $\Earth$ symbol indicates the position of the Earth where an adiabatic density ratio of 1.52 is predicted through the mantle (due to the phase changes in the transition zone
the observed density change in Earth's mantle is  rather 1.70).

This adiabatic density ratio controls the adiabatic temperature ratio according to \eqref{Ta} (see Figure \ref{Ratio}b). For the Earth, this ratio should be 
1.41 through the mantle (indicated by the symbol $\Earth$, say from 1600 K on top to 2256 K at the bottom). The maximum bottom temperature $T_a^b$ is, at any rate, bounded when $\rho_a \rightarrow \infty$ by
\eq{{T_a^b\over T_a^t} \leq e^{\Gamma_0}\approx 2.72. \eqlbl{Tbound}}
Even in very large silicated Super Earth, the bottom adiabatic temperature should remain moderate and hardly above a factor 2 times the surface adiabatic temperature
(Figure \ref{Ratio}b).

\subsection{The adiabatic temperature gradient}

According to \eqref{adiaq2}, the surface adiabatic gradient is simply 
\eq{\left.{dT_a\over d\tilde z}\right|_t = \Di T_a^t \eqlbl{dTa}}
where $\tilde z=z/H$ is the normalized height in the convective layer.
Obviously the heat carried out near the surface, along the adiabat, increases with $\Di$. However the adiabatic gradient
near the bottom is 
\eq{\left.{dT_a\over d\tilde z}\right|_b= \Di \left({\rho_a^t\over \rho_a^b}\right)^n  T_a^b}
In this expression, $T_a^b$ is bounded by equation \eqref{Tbound} and the thermal expansivity (related to the $\left({\rho_a^t / \rho_a^b}\right)^n$ term) decreases faster than $1/\Di$.
This means that the adiabatic gradient at depth (in absolute value), initially increases and then decreases with the dissipation number (inspection of \eqref{rhojump} shows that the adiabatic gradient
at depth decreases with $\Di^{-1/(n-1)}\approx \Di^{-0.43}$). This is visible in Figure \ref{figTa} (compare the adiabatic gradient at the bottom of the three curves). This implies that, as the dissipation number increases, the effects of compressibility become confined at shallow
depth, while deeper, the fluid appears more and more incompressible. Unexpectedly, when the effects of compression increase (when the planet radius increases), the deep convection appears more and more incompressible !

Another way to understand this is to consider that the compressible effects that affect convection are not related to $\Di$
which is based on the reference thermal expansivity but to $\overline \Di=\int_0^H \Di d\tilde z$, where the dissipation number \eqref{D} is averaged over the thickness of the layer. Using the expressions \eqref{alpha} and \eqref{rhoadia}, as shown in \citet{Ricard22}, one gets
\eq{\overline\Di \leq \Gamma_0,}
i.e., the average dissipation is never larger than the Gr\"uneisen parameter which is about 1.

\section{Convection in the mantle of super-Earths}

\subsection{Compressible convection}
In situations where the compressibility is important and where the physical parameters vary strongly with depth, the use of a simple Boussinesq convection model and the correction of the results by the a posteriori addition of an adiabatic
contribution are not sufficient. Using anelastic formulations \citep{oguraphillips, jarvis80, braginsky,lantz} may also be tricky, as it is easy to inadvertently contradict the basic thermodynamic rules \citep[see e.g.,][]{leng,albou13}. 
In a previous paper \citep{Ricard22}, we explained how we can solve the fully compressible equations without approximations, when inertia is neglected (the infinite Prandlt number
approximation), which is appropriate for mantle convection, i.e., how to solve
\begin{subequations}
\begin{align}
\Dt{\rho}+\rho \nablab \cdot \v{u}&=0, \eqlbl{FC:a}\\
\eta \nablab^2 \v u +{\eta\over 3} \nablab  \nablab \cdot \v u-\nablab { P}+\rho \v{g} \eqlbl{FC:b}&=0,\\
\rho T\Dt{\mathcal S}=\dot\varepsilon :\tau + &k \nabla^2 T, \eqlbl{FC:c}
\end{align}
\end{subequations}
where the viscosity $\eta$ and thermal conductivity $k$ are assumed uniform.
Following  exactly the rules of thermodynamics and starting from the EoS  \eqref{EoS1}, the
entropy can be expressed and by integration of $T{\rm d}\mathcal{S}=C_V {\rm d}T-\alpha K_T T  {\rm d}\rho/\rho^2$, writes
\eq{\mathcal{S}=C_V \ln {T\over T_a} + \alpha_0 K_T^0 \left({1\over \rho}-{1\over \rho_a}\right). \eqlbl{EntroND}}
which cancels out, at it should, when the density and temperature are those of the adiabatic conditions.
The adiabatic density and temperature are computed from \eqref{adiaq1} and \eqref{adiaq2}
\citep[see][for details]{Ricard22} where the heat capacity at constant pressure  that appears in the adiabatic profile is exactly given by Mayer's relation
\eq{C_P=C_V\left[1+\Gamma_0 \alpha_0 T\left( {\rho_0\over \rho} \right)^{n+1} \right].}

\subsection{Marginal stability and Schwarzschild criterion}

When heated from below, a fluid starts to convect when two conditions are required. First the local temperature gradient $|dT/dz|$ must be larger than the adiabatic gradient $|dT_a/dz|$. This is the Schwarzschild criterion
\citep{Schwarzschild1906}:
the temperature of a rapidly upwardly moving fluid  parcel follows the adiabatic gradient and must become warmer (e.g., less dense) that the surrounding to be gravitationally unstable. This criterion defines a necessary condition for convection. Second, the total temperature drop $\Delta T=T^b-T^t$ across the convective layer must be large enough so that a dimensionless number, the Rayleigh number, exceeds some critical value. In the simple case where the top and bottom boundaries are free slip, \citet{rayleigh} proved that his last condition can be expressed under the form
\eq{{\rm Ra}^c={\alpha_0 \rho_0^2 C_P g H^3 \Delta T \over \eta k} \geq {27\over 4} \pi^4=657.24 \eqlbl{Ra}}
This sufficient condition for convection was obtained in the Boussinesq approximation where all the parameters $\alpha_0$, $\rho_0$, $\eta$ and $k$ were uniform. In a compressible fluid $\alpha$ and $\rho$ are depth dependent and \eqref{Ra} is not necessary in agreement with the Schwarzschild criterion. As we are used to thinking that the adiabatic gradient is more or less uniform (this would be exactly true if the fluid were a perfect gas and which is roughly true in the Earth's mantle),  a superadiabatic Rayleigh number ${\rm Ra}_{sa}$ is generally defined where the temperature drop $\Delta T$ is replaced by the superadiabatic temperature drop $\Delta T_{sa}=\Delta T-\Delta T_a$, the temperature drop in excess of the adiabatic temperature drop
\eq{{\rm Ra}_{sa}={ \alpha_0\rho_0^2 C_P g H^3 \Delta T_{sa}\over \eta \kappa}={\rm Ra} {\Delta T_{sa}\over \Delta T},}
where $\alpha_0$ and $\rho_0$ are now some characteristic values of the depth-dependent thermal expansivity and density.
With this definition, ${\rm Ra}^c_{sa} \geq 657.24$ is in agreement with both what is found in the Boussinesq approximation and with the  Schwarzschild criterion \citep{malkus,GrLo01},
at least, if one assumes that $T_a$ varies linearly with depth (i.e., if one assumes that $dT_a/dz$ is uniform with $dT_a/dz=-\Delta T_a/H$). Various relations obtained in the Boussinesq approximation, for exemple between the heat flow and the Rayleigh number, are often considered to hold in the compressible case when the superadiabatic Rayleigh number is used.
We will show now that the situation is more complex in super-Earth cases where the adiabatic gradient is large and with a large curvature. 

To first discuss a simple case where the adiabatic gradient is constant, we consider a perfect gas with EoS
$P=\rho \mathcal{R} T$. This EoS is certainly not appropriate for a planetary mantle but it corresponds to the prototype case of compressible convection. For a perfect gas, 
$\alpha T=1$ and $C_P$ is constant in \eqref{adiaq2} so that $dT_a/dz=-\alpha_a T_a g/C_P=-g/C_P$. Using the surface temperature $T_0$ and the height $H$ of the convecting layer to non-dimensionalize the variable, one as simply $dT_a/dz=-\Di/\gamma$ where $\gamma=C_P/C_V$ is the heat capacity ratio (also known as adiabatic index or Laplace's coefficient). We then consider that a bottom temperature $rT_0$ is imposed, in which case the conductive geotherm is $dT/dz=-(r-1)$. The Schwarzschild criterion imposes therefore that convection cannot exist when $\Di \geq \gamma (r-1)$. 

\begin{figure}
\centerline{\includegraphics[width=8cm,angle=-90]{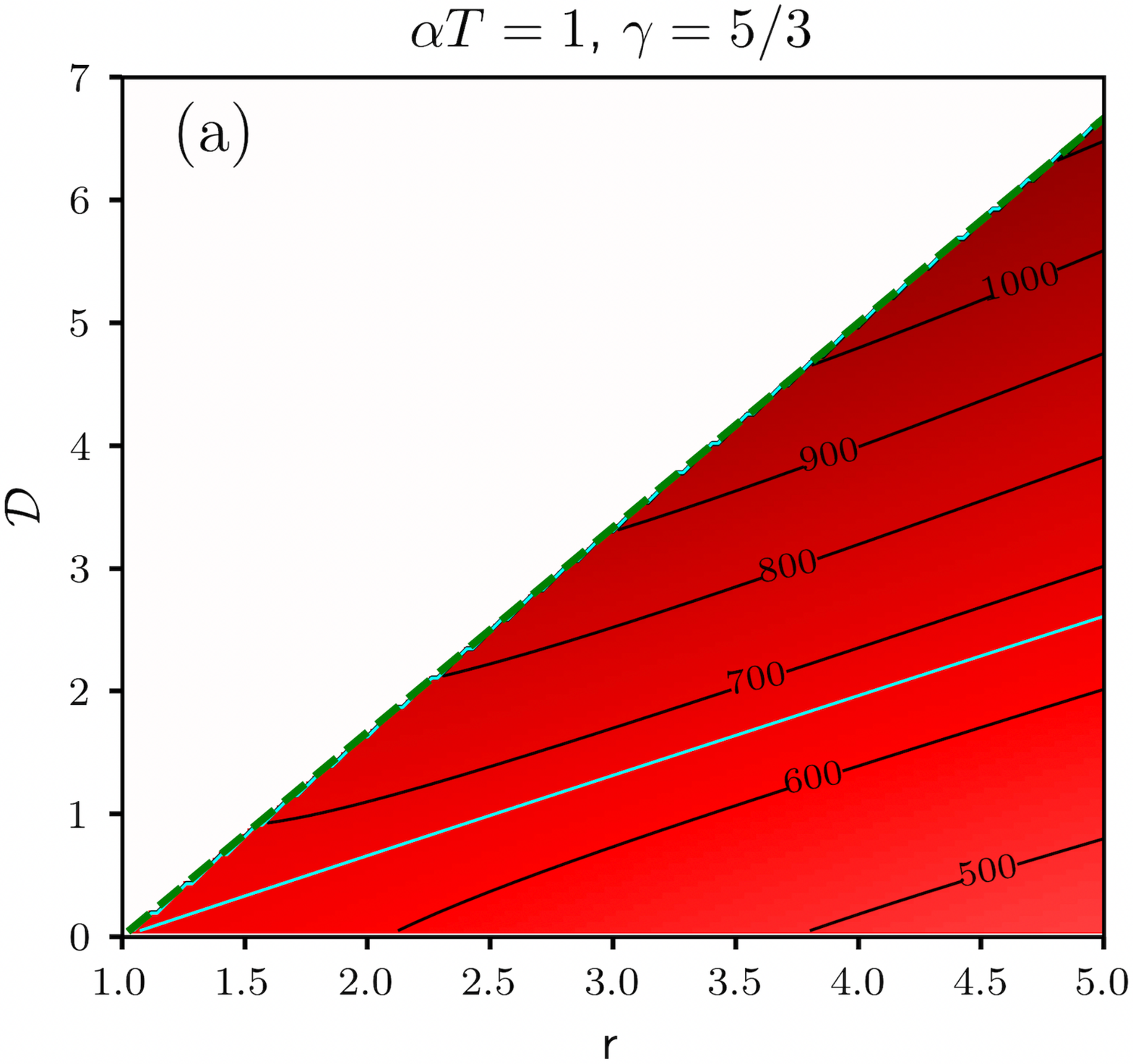} \hspace{-2cm} \includegraphics[width=8cm,angle=-90]{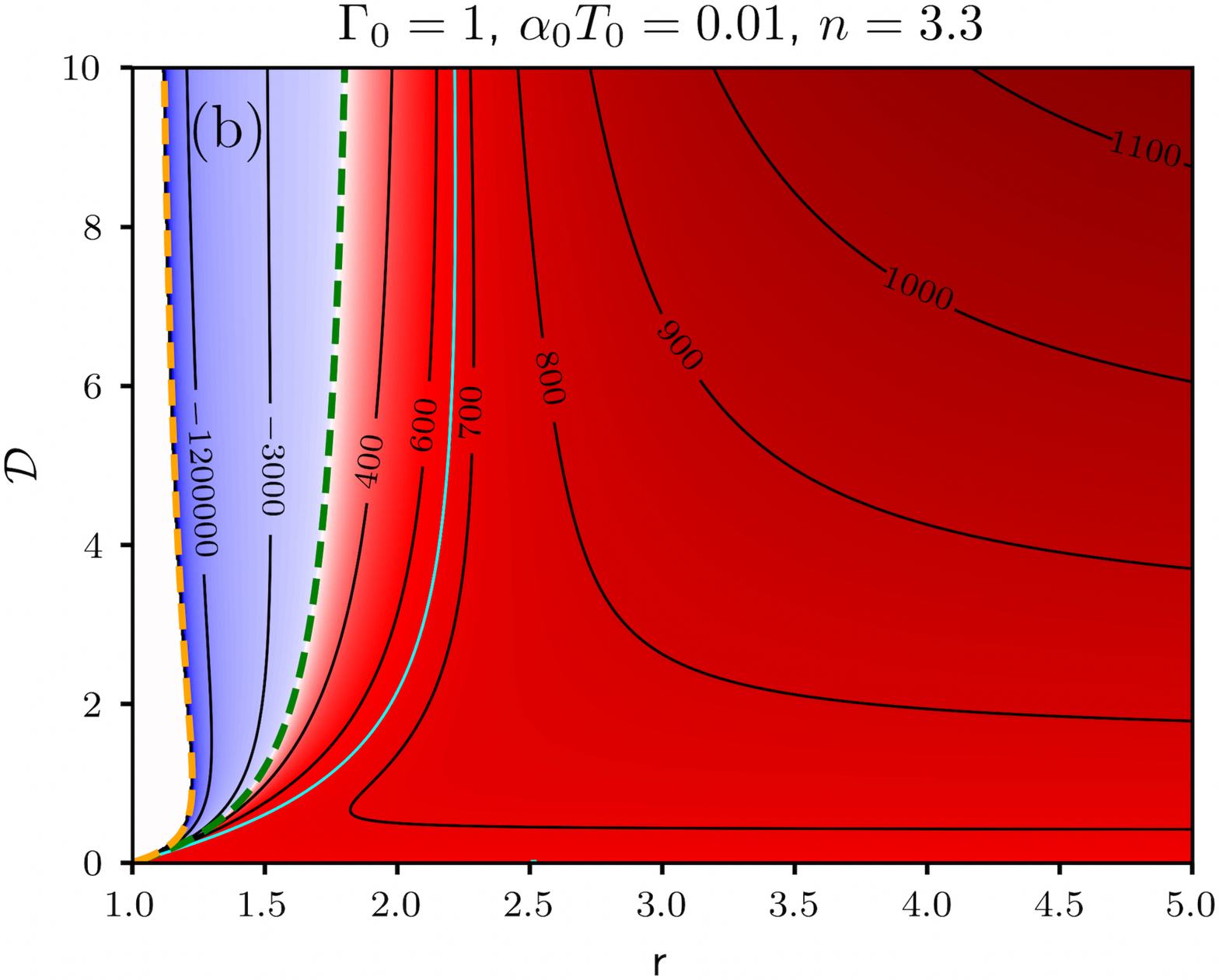}}
\caption{Critical superadiabatic Rayleigh number as a function of the surface dissipation number $\Di$, and the ratio between top and bottom temperatures $r$. In the left panel, the EoS of the convective fluid is that of a perfect gas(with $\gamma=5/3$),  in the right panel we consider the Murnaghan EoS, appropriated for condensed matter. While for the perfect gas, a large dissipation $\Di$ decreases the domain where convection can occur, for the Murnhagan solid, convection can easily start in the deep layers even when the imposed temperature difference is lower
than the adiabatic temperature difference. The dashed green and orange lines corresponds to the conductive profiles of Figure \ref{figTa}. The cyan lines in both panel are for ${\rm Ra}^c_{sa}=657.24$, the value obtained by Lord Rayleigh in the Boussinesq case.
}
\label{marginal}
\end{figure}

In a previous paper, we give the general equations verified by the marginally stable solution and how to compute the critical Rayleigh number for any EoS (equations 5.5-5.8 in \citet{albou17} and following comments). 
We therefore calculate the critical Rayleigh number ${\rm Ra}^c(\Di,r)$ for Rayleigh B\'enard convection of a perfect gas and plot the result in Figure \ref{marginal}a. 
As expected, convection can only occur below the $\Di = \gamma (r-1)$ line.
The cyan line corresponds to Ra$^c$=criterion
657.24, and indeed  for $\Di\rightarrow 0$ and $r\rightarrow 0$, the critical value obtained for the Boussinesq case is recovered. Increasing the temperature jump $r-1$ decreases the critical Rayleigh number, increasing the dissipation $\Di$ increases the critical Rayleigh number.

The situation is obviously very different on a planet where the adiabatic gradient appearing in the Schwarzschild criterion depends on the depth. Inspection of Figure \ref{figTa} for $\Di=10$ (green solid line), shows that convection cannot occur when the diffusive gradient $|dT/dz|$ is lower than that of the dashed orange line (tangent to the adiabatic profile at the bottom where the adiabatic gradient is minimal in absolute value). However until  $|dT/dz|$ reaches the value corresponding to the dark green dashed line (corresponding to $\Delta T=\Delta T_a$), convection can start in the deep layers
although $\Delta T \leq \Delta T_a$, i.e., although the superadiabatic Rayleigh number is negative. This is confirmed by the computation of the critical Rayleigh number shown in Figure \ref{marginal}b. There is a large domain in blue, where the conductive temperature gradient is in between that of the dark green curve and that of the orange curve of Figure \ref{figTa}, where convection can start in the deep layer while the critical superadiabatic Rayleigh number is negative.
Notice that again, when $\Di\rightarrow 0$ and $r\rightarrow 0$, the critical value computed for the Boussinesq case is recovered (${\rm Ra}^c_{sa} \rightarrow 657.24$, a value shown by a cyan line). This limit is indeed independent of the chosen EoS.

\subsection{Convection simulations}

With the same numerical code as in \citet{Ricard22}, we solve the mass, momentum and energy conservation system \eqref{FC:a}-\eqref{FC:b}-\eqref{FC:c}
with the software Dedalus \citep{dedalus}, which handles coupled differential equations that are solved iteratively using a spectral decomposition. 
In our simulations, the surface temperature is $T_t=T_0$ and the bottom temperature $T_b=r T_0$. We assume that the surface pressure is $P_t=P_0=0$ (hence that the surface density is $\rho_0$). 

As discussed in \citet{Ricard22}, it is tricky to work with a compressible fluid on a fixed numerical grid. Indeed, we do not know what the initial mass in the convective layer (defined by an initial density profile) must be to ensure that when convection is well established, the surface pressure is zero.  We therefore perform our simulations for given ${\rm Ra}$, $r$ and $\Di$, starting with different initial masses (different initial assumptions of the density profile) until the average surface pressure is statistically zero (the local surface pressure itself remains a function of space and/or time and is commonly interpreted as equivalent to express the presence of a dynamic topography induced by convection). 

\subsection{Heat flux in compressible convection}

For compressible convection at infinite ${\rm Pr}$ number, the superadiabatic heat flux (the surface heat flux $Q$ minus the surface adiabatic heat flux $Q_a$) and the superadiabatic Rayleigh number are related by
\eq{{\rm Nu}_{sa}={Q-Q_a\over \Delta T-\Delta T_a}  \propto {\rm Ra}^{1/3}_{sa}, \eqlbl{RaNu}}
where ${\rm Nu}_{sa}$ is the superadiabatic Nusselt number (in equation  \eqref{RaNu}, the dimensionless heat flows are normalized by $kT_0/H$ and the temperatures by $T_0$). This scaling law with the superadiabatic Rayleigh number is
in agreement with what is found in the Boussinesq approximation \citep{malkus,GrLo01}. This expression is usually proposed in situations where $\Delta T_a$ is smaller
and often much smaller than $\Delta T$ which is not necessarily verified  when $\Di$ is large,  as convection can occur even with $\Delta T_a \geq \Delta T$.
Notice also that, as shown by the blue dashed line in Figure \ref{figTa}, the adiabatic heat flow at the surface may be very large and $Q-Q_a$ may be negative even in the case where  $\Delta T_a\leq\Delta T$. The "adiabatic" Nusselt number, ${\rm Nu}_a=Q_a/\Delta T_a=\Di T_a^t/\Delta T_a$ (see \eqref{dTa}) is of order $\Di$
as $T_a^t/\Delta T_a\approx 1$. For $\Di=10$, this is already a large heat flow which requires a Rayleigh number of 10$^5-10^6$ in the Boussinesq case.

This situation where the convective heat flow can be lower than the adiabatic heat flow, is not unknown and may actually happen in the Earth's core (although inertial, electromagnetic and rotationnal effects not accounted for in our model become crucial in the core). The conductivity of iron is large enough \citep{staceyloper,dekoker,gomi} that in the top part of the core, the heat transported along the adiabat may be larger than that carried out by convection \citep{labrossepoirierlemouel,listerbuffett}. This would imply the presence of a stratified layer in which the non-adiabatic temperature
carries the heat flow downward to balance the upward transport along the adiabat. How the deep convection interacts with a shallow layer with a large adiabatic gradient can now be discussed with a few numerical simulations. 
There is a very large number of quantities of interest that could be computed from these numerical simulations. Here, we will simply discuss the general pattern of convection and the average temperature profiles, and how the energy is transported through 
the whole layer. 

In the Boussinesq approximation used as a reference model, the heat flow through the fluid is simply
\eq{Q_{Bo}= \overline{\rho w C_V T}-k{d\overline{T}\over dz},\eqlbl{Qbous}}
where $w$ is the vertical velocity and $T$ the total temperature. The overbar indicates here an average of the various quantities horizontally and in time; this heat flow is constant with depth. In the Boussinesq approximation, the density $\rho$ is a constant, just as the heat capacity $C_V$ and the heat capacity at constant volume or constant pressure are not distinguished. 
In a compressible fluid  \citep[see][for details]{Ricard22}, the relevant quantity advected by the flow is the enthalpy $\mathcal{H}$ 
that can be deduced from \eqref{EoS1} by integration of ${\rm d}\mathcal{H}=T{\rm d}\mathcal{S}+{\rm d}P/\rho$
\eq{\mathcal{H}=\left( C_V   + {\alpha_0 K_T^0\over \rho}  \right)T +{ K_T^0\over (n-1)} {\rho^{n-1} \over \rho_0^{n}},  \eqlbl{EnthalND}}
and the heat flow can then be written as 
\eq{Q=\overline{\rho w C_V  T_{sa}}-k{d\overline{T_{sa}}\over dz}+\overline{\rho w (\mathcal{H}-C_VT_{sa})}-k{dT_a\over dz}-\overline{u\tau_{xz}+w\tau_{zz}}, \eqlbl{Q}}
where we consider separately the adiabatic $T_a$ and superadiabatic $T_{sa}=T-T_a$ temperature.

The first two terms converge to their counterparts  of \eqref{Qbous} when the adiabatic temperature is a constant, the third term (which can obviously be simplified with the first one) is a correction due
to the fact that the enthalpy rather than the specific heat $C_V T$ is transported. The fourth term is the conduction along the adiabat and the last one is the work flow ($\tau_{xz}$ and $\tau_{zz}$
are the deviatoric stresses, $u$ the horizontal velocity). Had we used dimensionless variables, then the ratio between this term and the first one would be $\Di \Delta T/ ({\rm Ra} T_0)$ and indeed, would become negligible in the Boussinesq approximation when $\Di \rightarrow 0$.

\subsection{Convection at small dissipation and large temperature drop ($\Di=1$, $r=10$)}

\begin{figure}
\centerline{\includegraphics[width=18cm,angle=0]{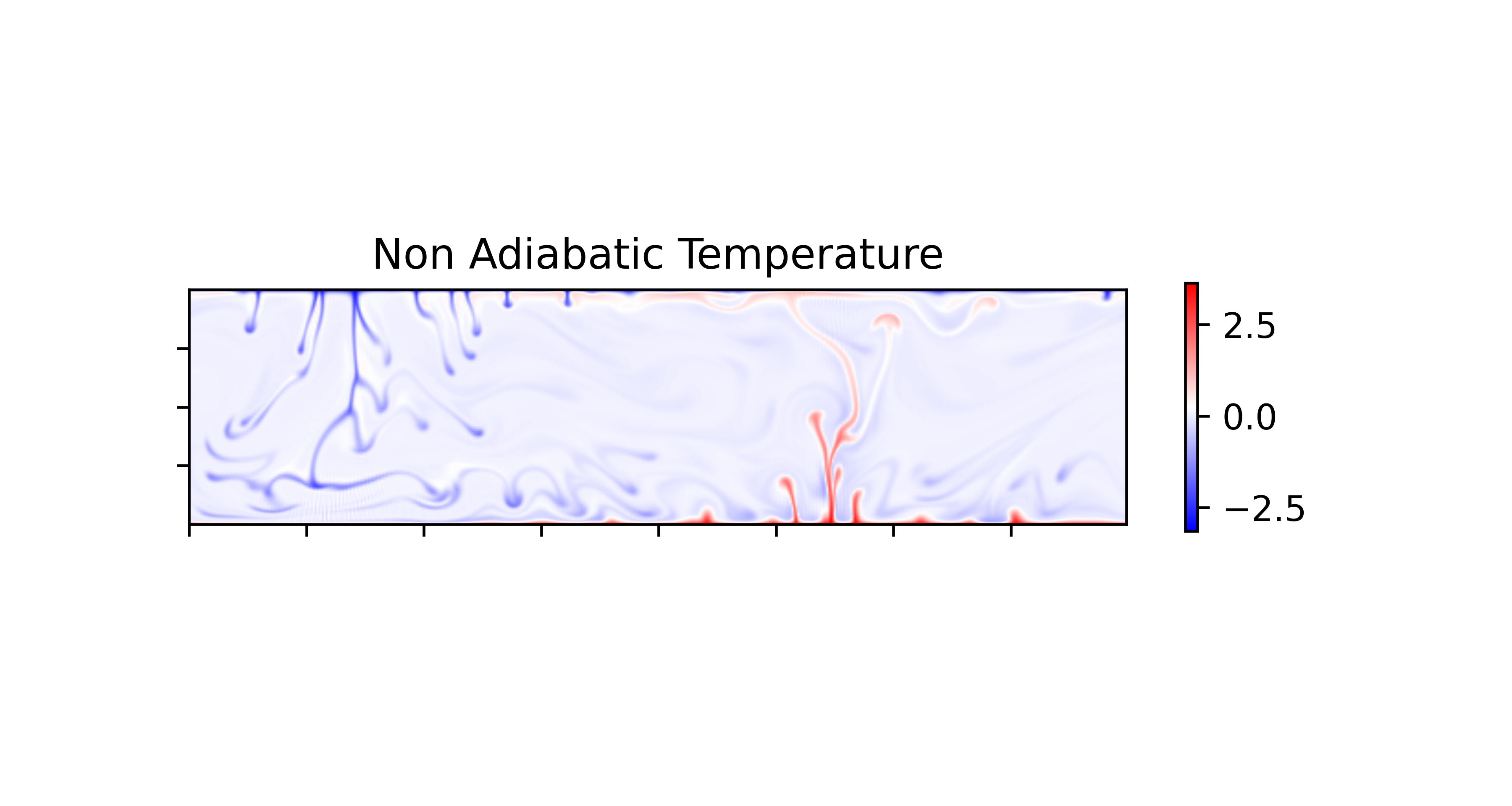}}
\vskip -2cm
\caption{
Snapshot of the superadiabatic temperature in the convective layer with ${\rm Ra}_{sa}=10^8$, $\Di=1$ and $r=10$. The downwellings and upwellings are less stable and less continuous with depth than in the Boussinesq approximation.
The plumes are not homogeneously distributed but tend to form clusters. 
}
\label{D1r10-1}
\end{figure}

In a first simulation at moderate Rayleigh number ${\rm Ra}_{sa}=10^8$, we consider a situation in which the adiabatic temperature drop 
is small compared to the imposed temperature difference, with $\Di=1$ and $r=10$ which corresponds  more or less to terrestrial conditions. A typical snapshot of the non-adiabatic temperature field is depicted in Figure \ref{D1r10-1}. 
This situation remains relatively close to the classical Rayleigh-B\'enard  convection in the Boussinesq approximation but some differences are notable. Due to dissipation, the descending and ascending plumes are rather discontinuous. They tend to form clusters, which is consistent with the suggestion by \citet{schubert04} that the two superplume regions observed in the deep Earth's mantle may be clusters of smaller
plumes whose heads have merged into a large region of hot and buoyant material.

\begin{figure}
\centerline{\includegraphics[width=9cm,angle=0]{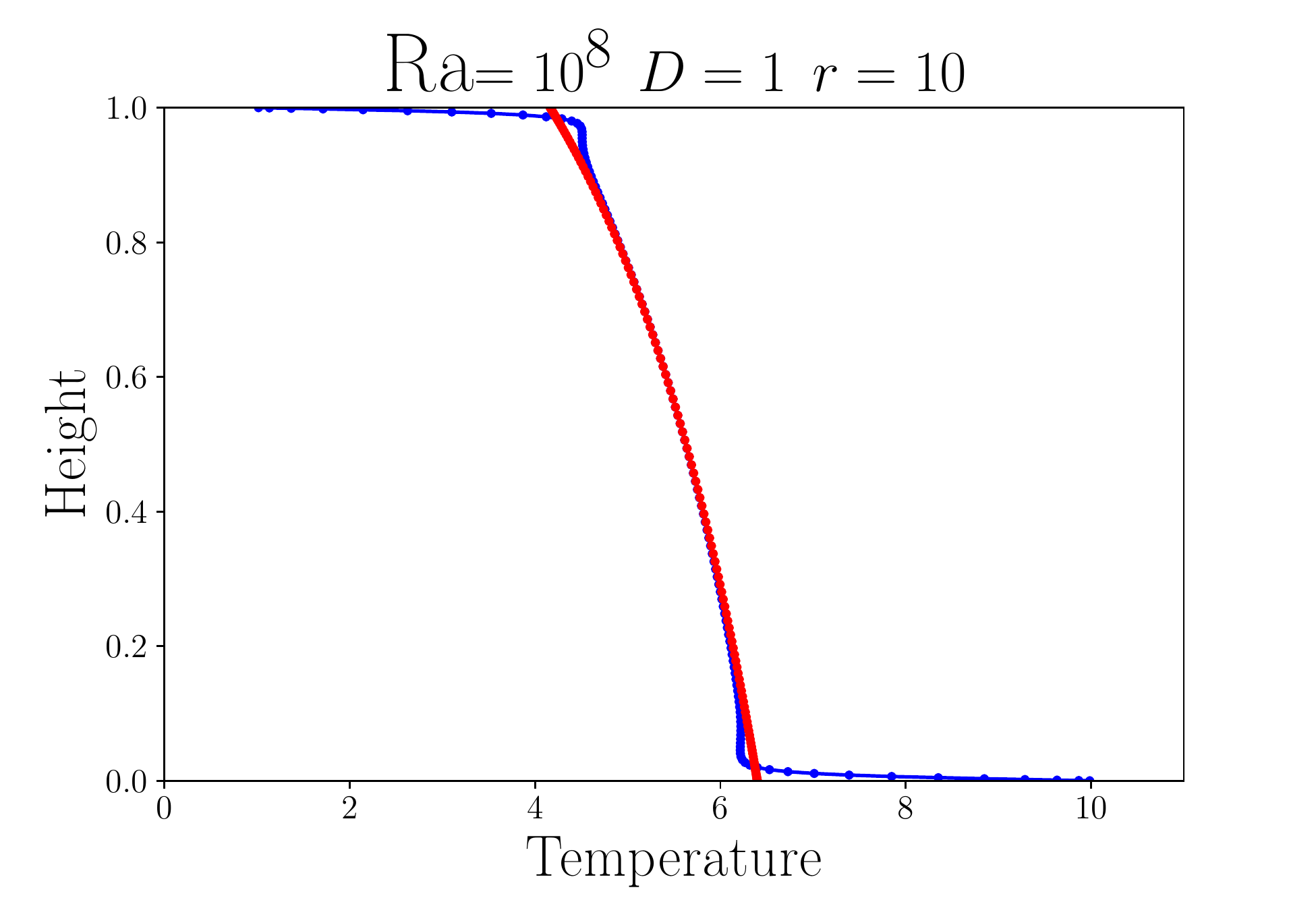}}
\caption{
Temperature profile (blue) and adiabatic (red) in a convection model with ${\rm Ra}_{sa}=10^8$, $\Di=1$ and $r=10$. The mass of the fluid that can be computed from the adiabatic and hydostatic profile is by construction the total mass of the fluid. The pressure average at the surface and over time is zero. The two boundary layers have similar thicknesses. The temperature overshot near the hot bottom boundary layer is slightly more pronounced that that under the lithosphere. The total temperature jump across the layer is $r-1=9$ and the adiabatic temperature jump is 2.22.
}
\label{D1r10-2}
\end{figure}

The time-average temperature profile is depicted in blue in Figure \ref{D1r10-2}. The dots along the temperature profile correspond to the nodes of the Chebyshev polynomials used by the software Dedalus. 
Top and bottom boundary layers have comparable thicknesses. The adiabatic profile (red) is quasi linear and provides a close approximation of the real temperature (Figure \ref{D1r10-2}). The total mass under the actual temperature profile and under the adiabatic profile is the same and leads to a zero average pressure at the surface when the convection is statistically steady.


\begin{figure}
\centerline{\includegraphics[width=7.5cm,angle=0]{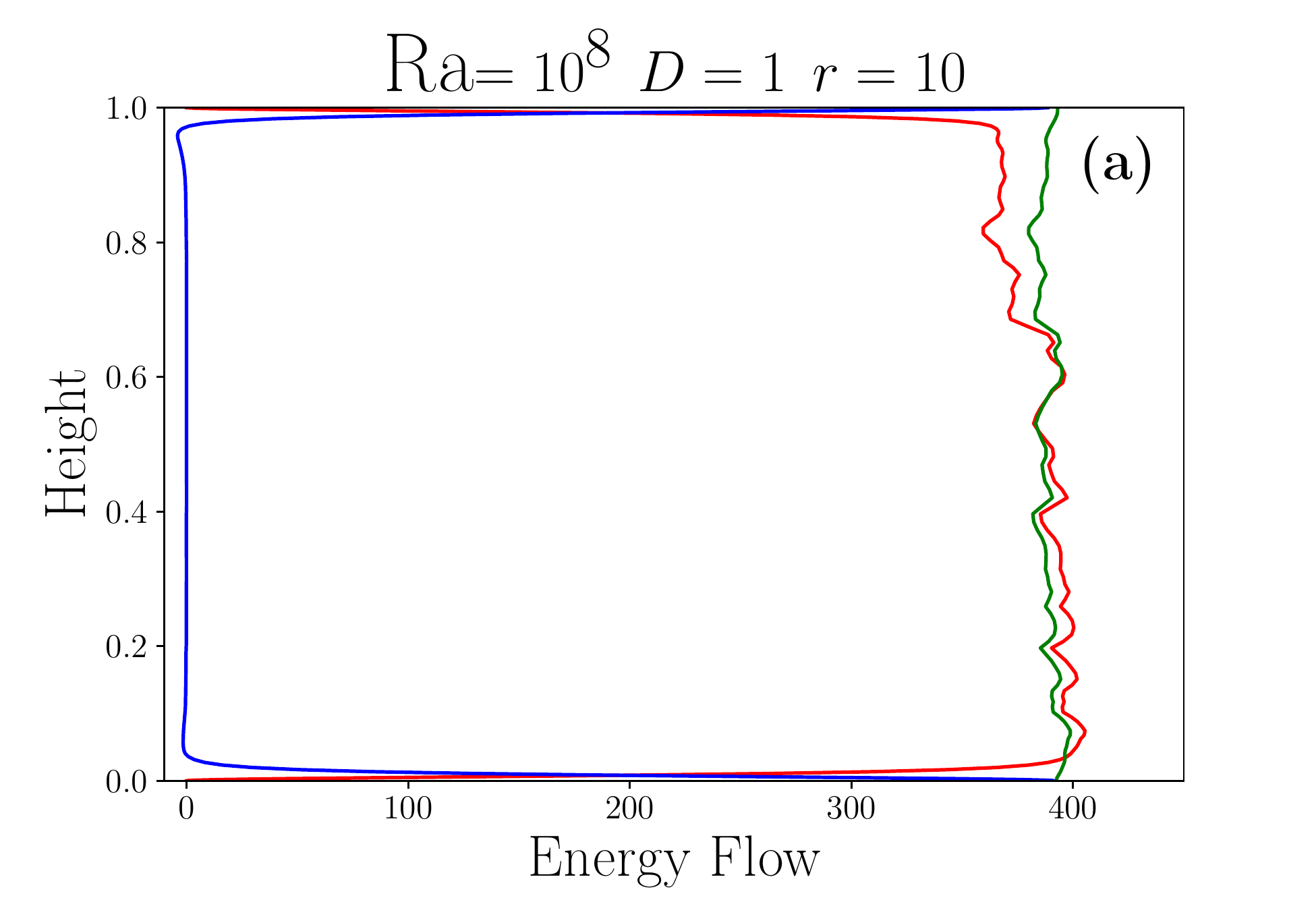}\includegraphics[width=7.5cm,angle=0]{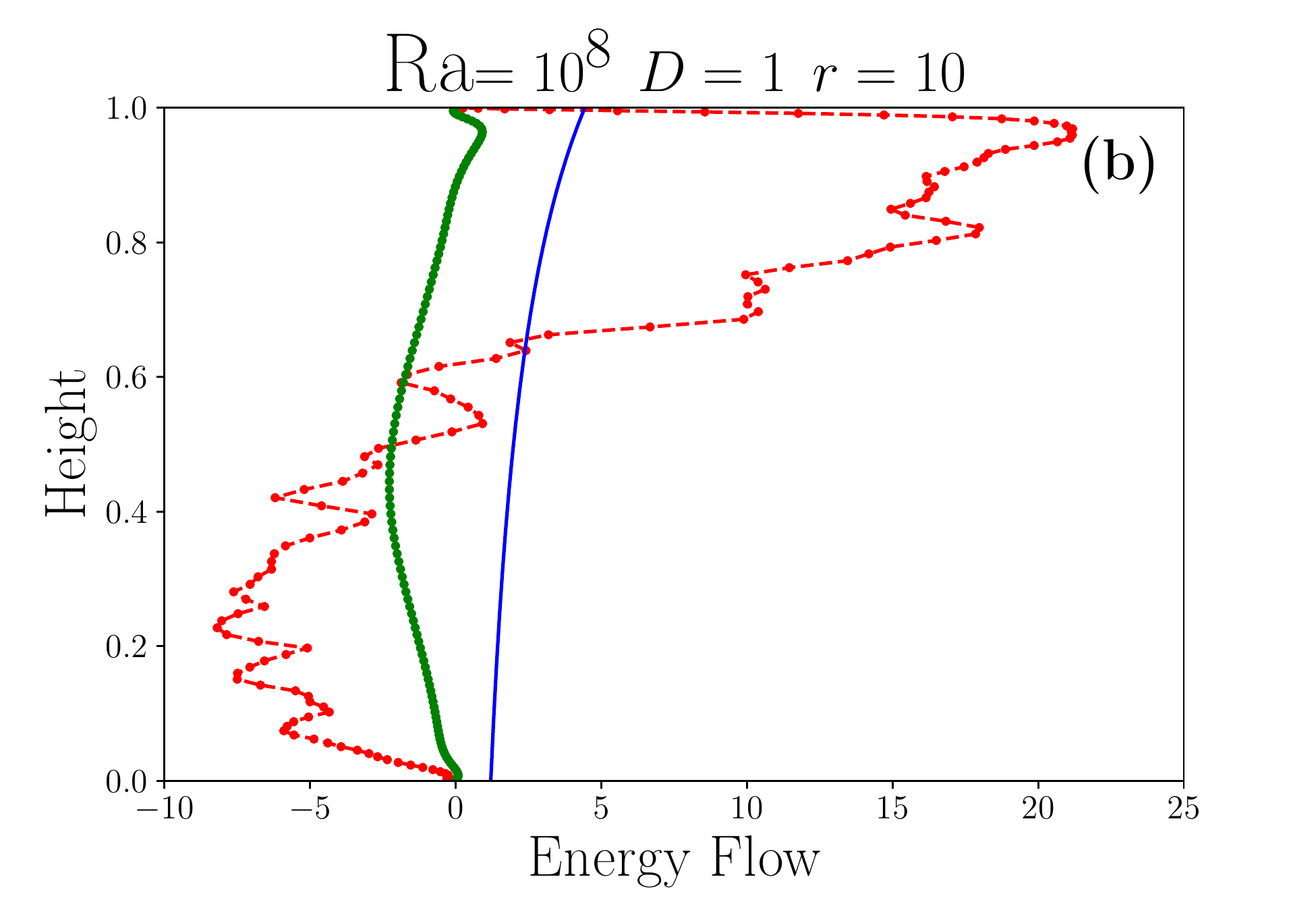}}
\caption{
Energy flow across the convective layer. In panel (a), we plot the total energy flow, averaged on time, in green. This flux should become independent of depth if we had performed the averaging over an infinitely long time window. The two major contributors of the energy are the advection of specific heat  (red) and the conduction along the non-adiabatic thermal profile (blue). The former is efficient in the bulk of the fluid, the latter in the boundary layers. The other minor contributors of energy transport (panel (b)) are the conduction along the adiabat (blue), the work flow (green), the contribution due to the fact that enthalpy rather than specific heat is transported (red). Notice the difference in scale between the two panels.
}
\label{D1r10-3}
\end{figure}

In this simulation the time averaged heat flow is $Q=398$, $Q_a=4.41$, $\Delta T_{sa}=6.78$ and the Nusselt number, ${\rm Nu}_{sa}=27.37$ and therefore ${\rm Nu}_{sa}=0.13\,{\rm Ra}_{sa}^{1/3}$. The prefactor is in agreement with other simulations performed in the Boussinesq regime \citep{sotinlabrosse} or in the fully compressible case with an ideal gas Eos \citep{jezabel}. We depict in Figure \ref{D1r10-3} the profiles of the various components of the heat flow (see \eqref{Q}). In the left panel (a), we plot the total heat flow (green), the transport of specific heat (red) and the conduction along the non-adiabatic temperature profile (blue) (i.e., the total heat flow $Q$ and the two first terms of equation \eqref{Q}). The total heat flow (green) would be depth-independent when averaged over a very long time. The red and blue components (the specific heat transport and the conductive term)
are the only terms carrying heat in the Boussinesq approximation (see \eqref{Qbous}). Due to compressibility, other minor contributors to the energy transport are present in panel \ref{D1r10-3}b. The transport of $\mathcal{H}-C_VT_{sa}$ (red), the conduction along the adiabat (blue), and the work flow (green) (i.e., the third, fourth and fifth terms of \eqref{Q}).
These different terms tend to increase the energy transport near the surface and decrease it at depth. The compressibility does not affect the energy transport very much as $\Di$
remains small compared to the applied total temperature difference across the layer and the minor components of panel b have amplitude at most of $\approx 5\%$ of the major components of panel a.

\subsection{Convection at large dissipation and large temperature drop ($\Di=r=10$)}
\label{10-10}

\begin{figure}
\vskip -5cm
\centerline{\includegraphics[width=18cm,angle=0]{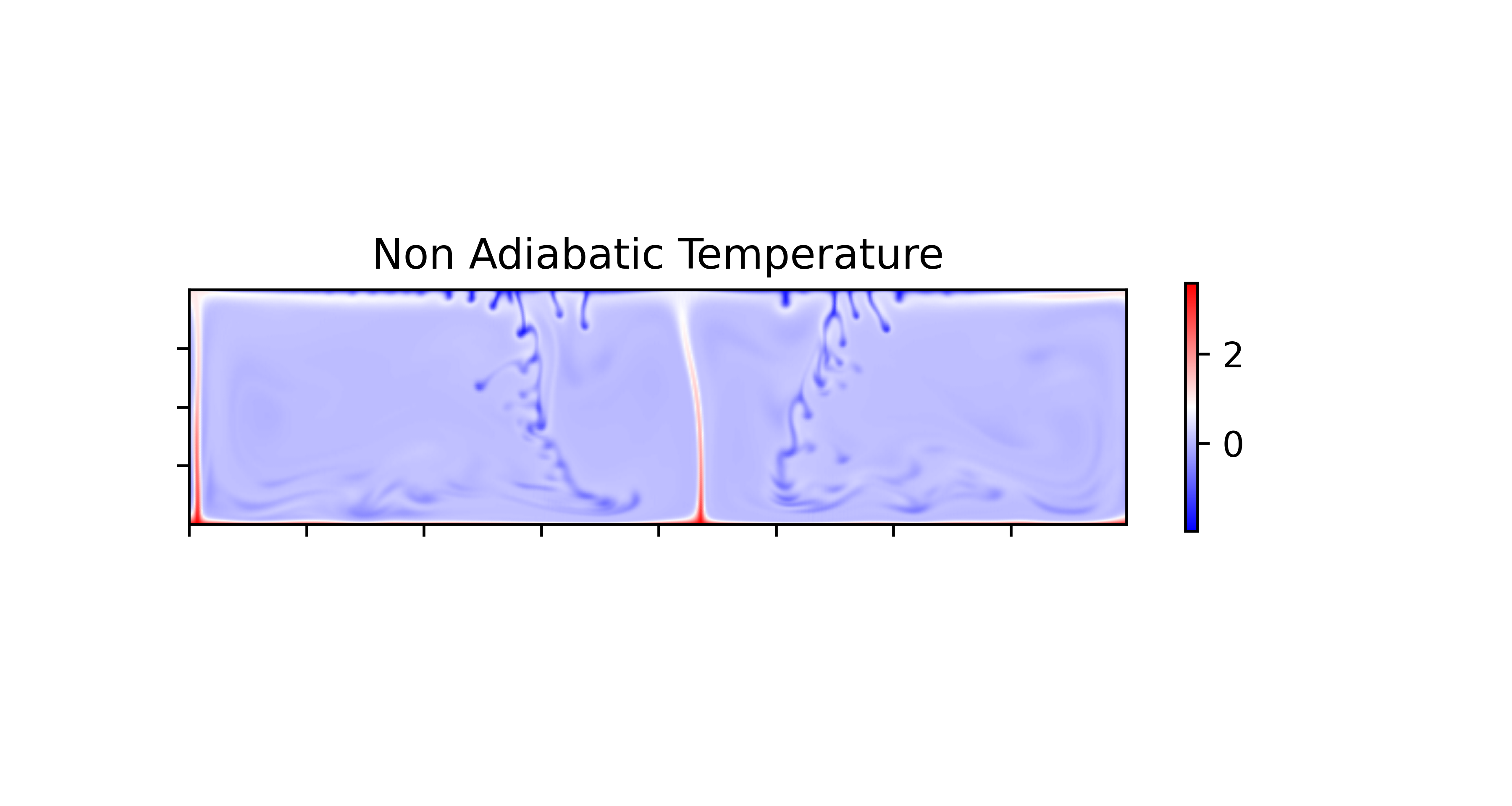}}
\vskip -2cm
\caption{
Snapshot of the superadiabatic temperature in the convective layer with ${\rm Ra}_{sa}=10^8$, $\Di=10$ and $r=10$. The downwellings are still less stable and less continuous with depth than the hot plumes but for both hot and cold plumes, crossing the mantle becomes difficult.
}
\label{D10r10-1}
\end{figure}

\begin{figure}
\centerline{\includegraphics[width=9cm,angle=0]{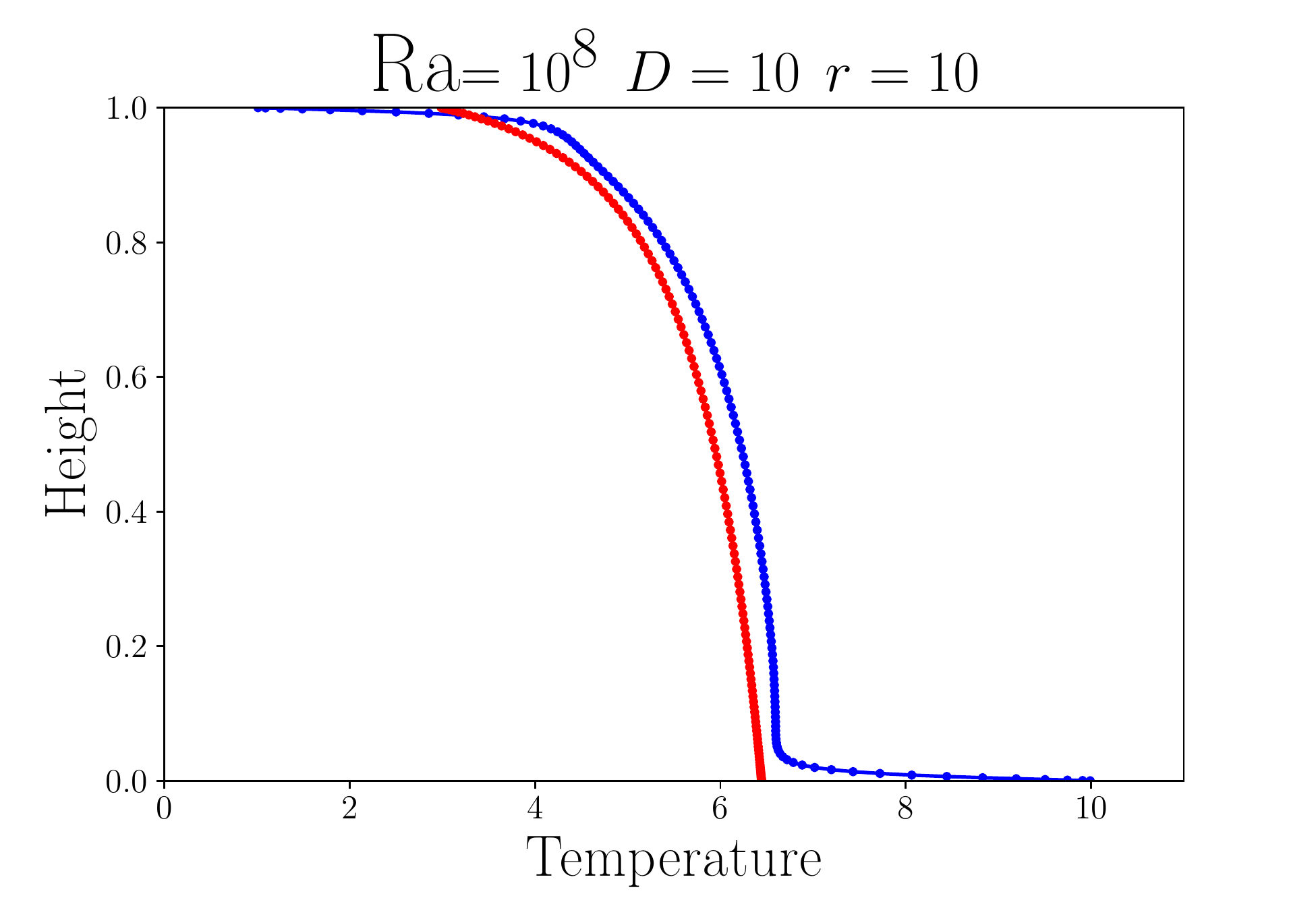}}
\caption{
Temperature profile (blue) and adiabatic (red) in a convection model with ${\rm Ra}=10^8$, $\Di=10$ and $r=10$. The mass of the fluid that can be computed from the adiabatic profile is by construction the total mass of the fluid. The total average pressure at the surface, average over time is zero. The thickness of the top boundary is strongly affected by the ability of the fluid to carry heat along the adiabat.
}
\label{D10r10-2}
\end{figure}

We can now consider the case where the dissipation number becomes comparable to the temperature ratio, $\Di=r=10$. A typical snapshot of the non-adiabatic temperature is depicted in Figure \ref{D10r10-1}. At large dissipation number the cold plumes can hardly cross continuously the convective layer \citep{albou22}. As already observed in \citet{hansen93}, the hot instabilities gain buoyancy while they rise in the mantle as the thermal expansivity increases. They are stronger
and more stationary than the cold instabilities that lose buoyancy with depth. The temperature profiles are shown in Figure \ref{D10r10-2}. Due to the large heat flow carried out along the adiabat, the cold boundary layer (the lithosphere) is poorly defined and its thickness increases compared to the Boussinesq case. On the contrary the hot deep boundary layer is much less affected as the adiabatic gradient is here minimal. The adiabatic profile (red) and the actual temperature (blue) have roughly similar curvatures. We recall that both thermal profiles correspond to the same total mass in the mantle ; the red adiabatic curve has not been computed to provide a best fit to the observed temperature. Furthermore, the idea that the actual temperature profile should be adiabatic is based on the assumption that the dissipation is negligible which is not the case when $\Di$ is large.

In this simulation the heat flow is $Q=270$, $Q_a=30.75$, $\Delta T_{sa}=5.56$ and the Nusselt number, ${\rm Nu}_{sa}=43.03$ and therefore ${\rm Nu}_{sa}=0.09 \,{\rm Ra}_{sa}^{1/3}$. The prefactor of this equation is smaller than usually found:  the Nusselt number, which is inversely related to the thickness of the cold boundary layer, is small because this thickness is increased by the large flux carried by conduction along the adiabat.

\begin{figure}
\centerline{\includegraphics[width=7.5cm,angle=0]{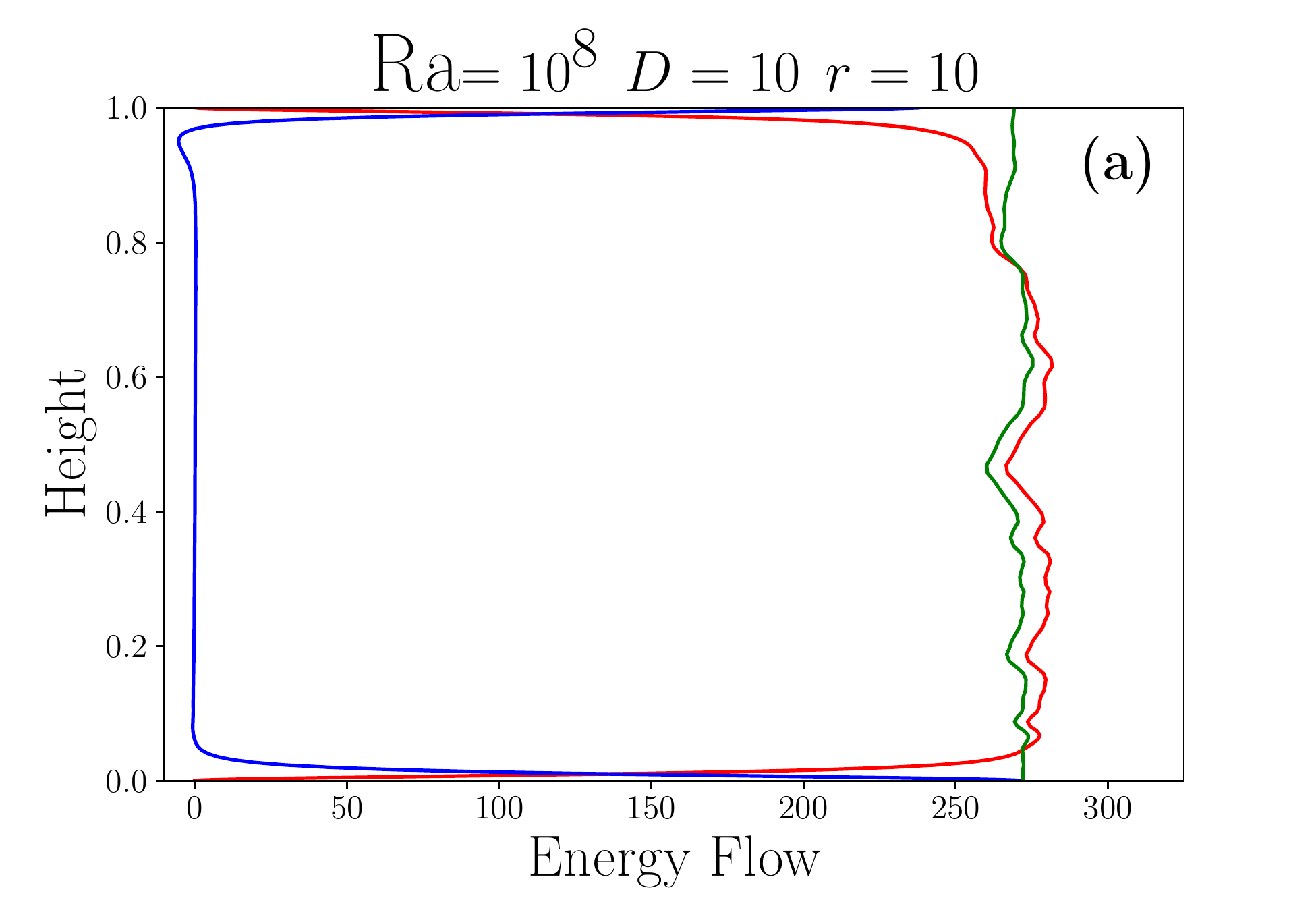}\includegraphics[width=7.5cm,angle=0]{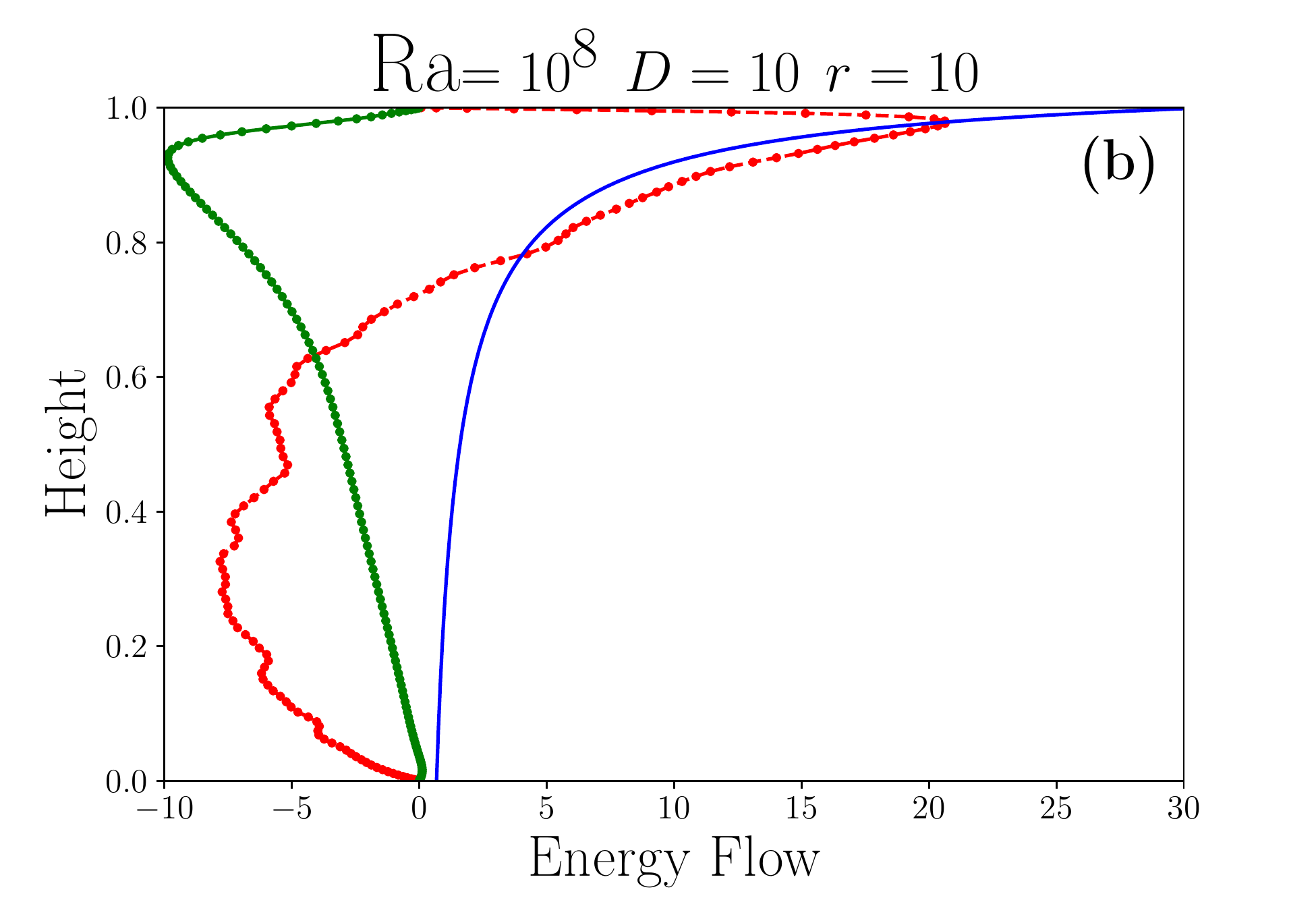}}
\caption{
Energy flow across the convective layer. In panel (a), we plot the total energy flow, averaged on time, in green. The other minor contributors of energy transport are the conduction along the adiabat (blue, panel (b)), the work flow (green, panel (b)), the contribution due to the fact the enthalpy transport is not restricted to the $C_V T$ term (red, panel (b)). Notice the difference in scale between the two panels. 
}
\label{D10r10-3}
\end{figure}

Similarly to Figure \ref{D1r10-3}, the various components of the heat flow are depicted in Figure \ref{D10r10-3} when the dissipation is now $\Di=10$. The transport of specific heat  (red, panel a), like for the previous case, underestimates the energy transport in the upper part of the mantle and overestimates the energy transport in the lower part. The heat conducted along the non-adiabatic gradient (blue, panel (a)) is lower across the top cold boundary where a significant heat (12\% of the surface heat flow) is carried out along the adiabat (blue curve, panel (b)). The energy transfert due to the difference between enthalpy and specific heat (red, panel (b)) and the work flow (green, panel (b)) are also significant. These three minor components added to the two major components of panel (a), lead to a global heat flow (green, panel (a)), independent of depth.

\subsection{Convection at large dissipation and small temperature drop ($\Di=10$ and $r=2$)}
\label{10-2}
\begin{figure}
\centerline{\includegraphics[width=16cm,angle=0]{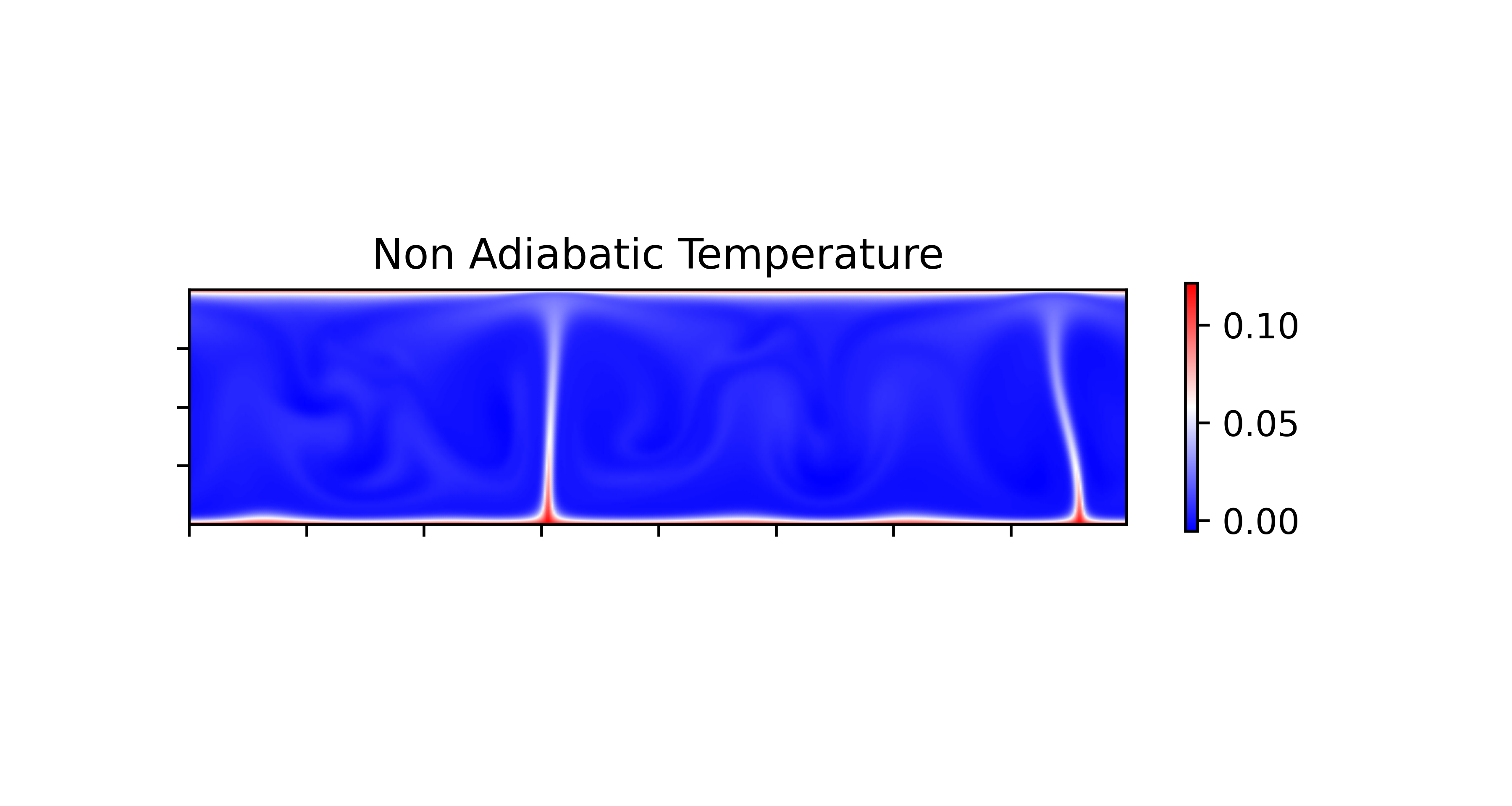}}
\vskip -2cm
\caption{
Snapshot of the superadiabatic temperature in the convective layer ($\Di=10$ and $r=2$). The hot rise plumes spread under an even hotter (in terms of non-adiabatic temperatures)
top boundary layer.
}
\label{D10r2-2}
\end{figure}

\begin{figure}
\centerline{\includegraphics[width=9cm,angle=0]{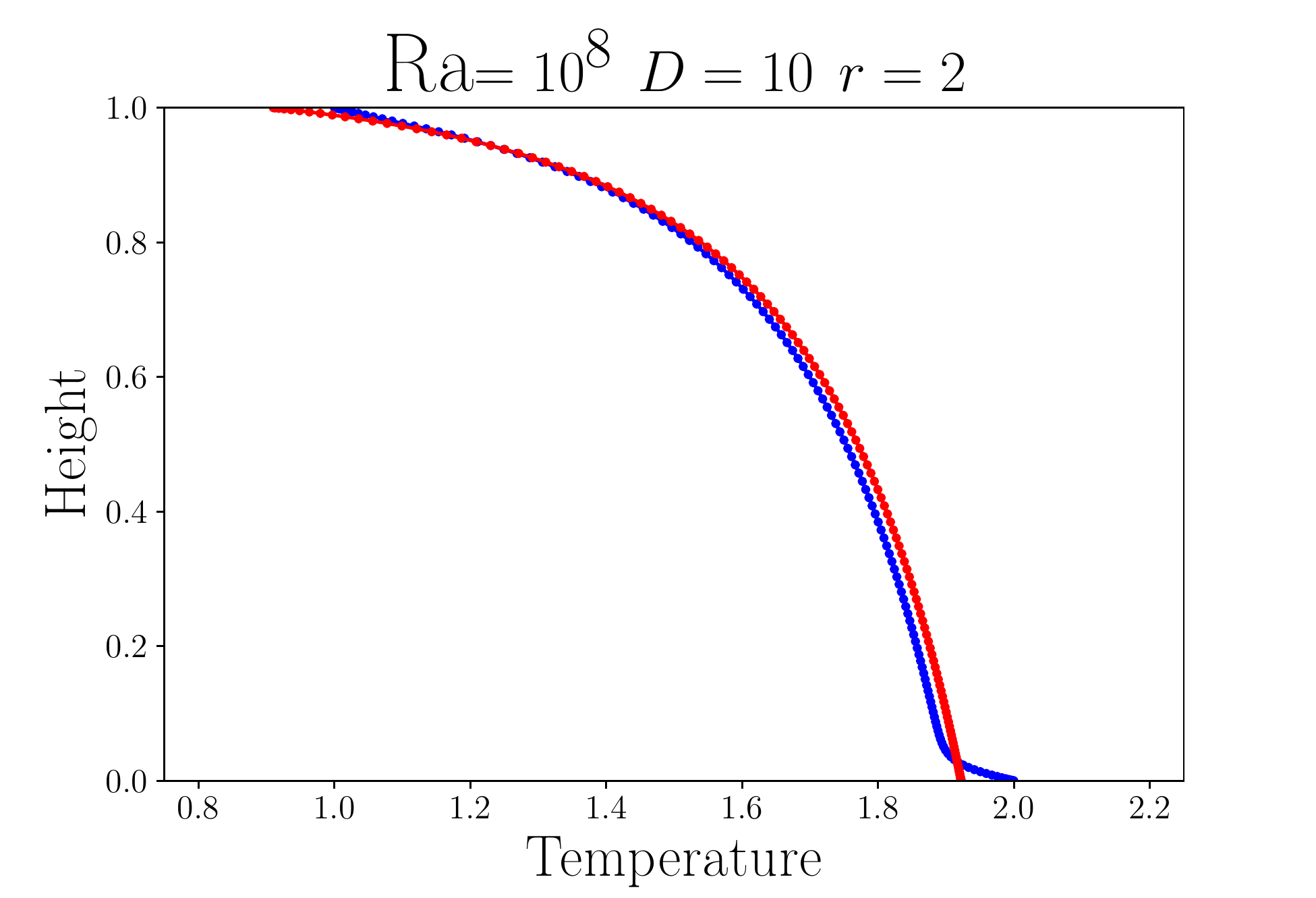}}
\caption{
Temperature profile (blue) and adiabatic (red) in a convection model with ${\rm Ra}=10^8$, $\Di=10$ and $r=2$. Except for a bottom boundary layer the non-adiabatic average temperature (blue) follows the general curvature of the adiabat (red).}
\label{D10r2-1}
\end{figure}

As we discussed, convection can also occur with a negligible superadiabatic temperature jump $\Delta T_{sa}$.  We therefore perform a simulation, always at ${\rm Ra}_{sa}=10^8$ but with $\Di=10$ and $r=2$. In that case  $\Delta T_{sa}=0.01$. A temperature snapshot is shown in Figure \ref{D10r2-1}. Rising plumes are strong and vigorous but no downwelling currents are visible. In fact,  no cold boundary exist and on the contrary the surface corresponds to a maximum of superadiabatic temperature. These characteristics are also obvious in Figure \ref{D10r2-2} depicting the average temperature profile (blue). The adiabatic temperature (red) has a lower surface temperature that the real surface temperature.

The various components of the energy transport are shown in Figure \ref{D10r2-2}. The balance between the various terms is very different from the previous cases (see Figures \ref{D1r10-2} and \ref{D10r10-2}). In this simulation the heat flow is $Q=4$, $Q_a=8.9$, and the superadiabatic Nusselt number as defined in \eqref{RaNu}  would be negative ${\rm Nu}_{sa}=-490$. When the adiabatic gradient has a large curvature with depth, the usual {\rm Ra-Nu} relation cannot be used. Instead of transfering heat to the surface, conduction along the superadiabatic temperature profile (panel a, blue) drives the energy downward. The advection of specific heat (red panel a) transports the energy in the deep convective layer, but the main transport occurs at shallow depth, along the adiabat (blue, panel b). The work flow (green) appears negligible and the distinction between enthalpy and specific heat (red, panel b) only transports a small excess of energy. 

\begin{figure}
\centerline{\includegraphics[width=7.5cm,angle=0]{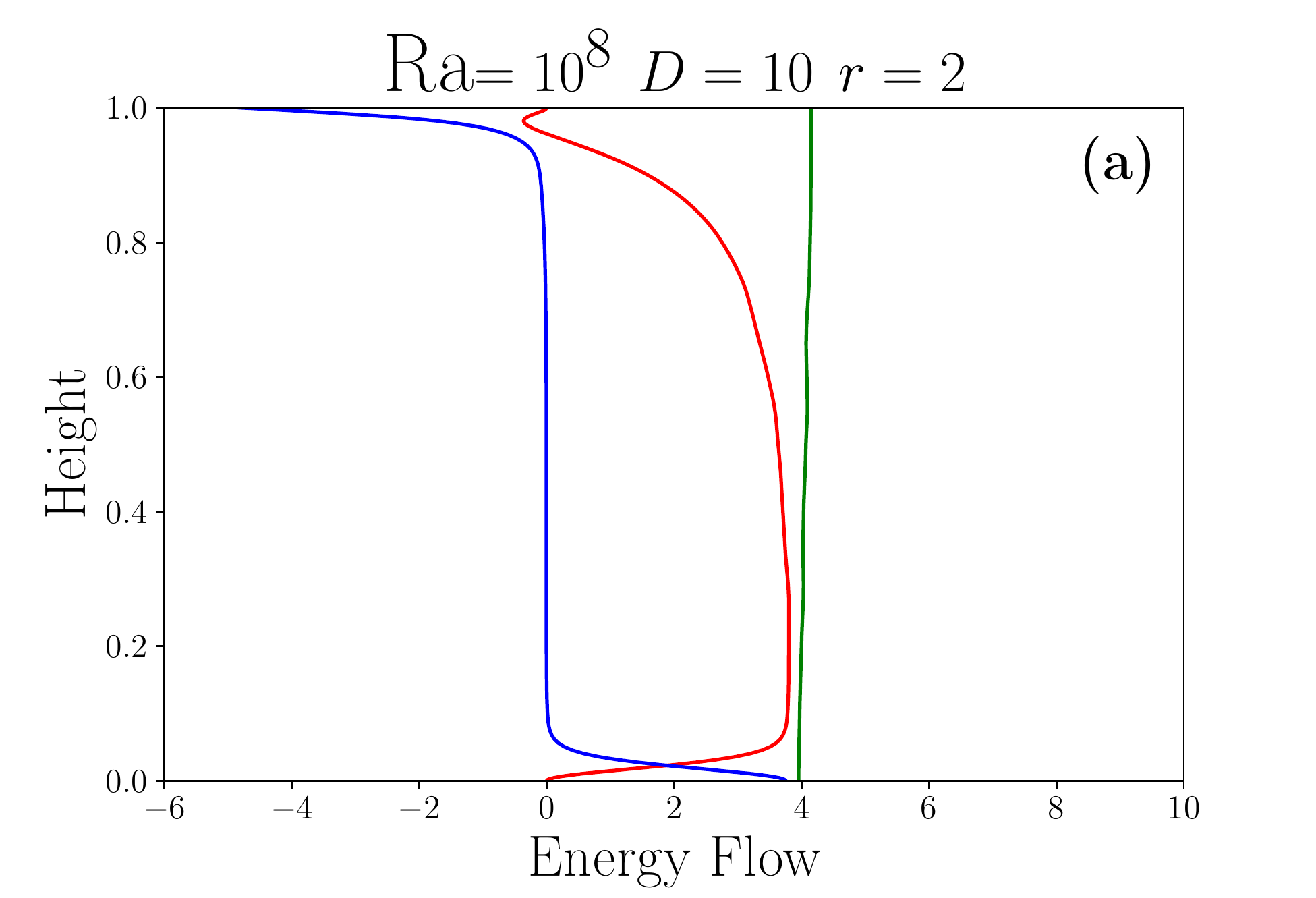}\includegraphics[width=7.5cm,angle=0]{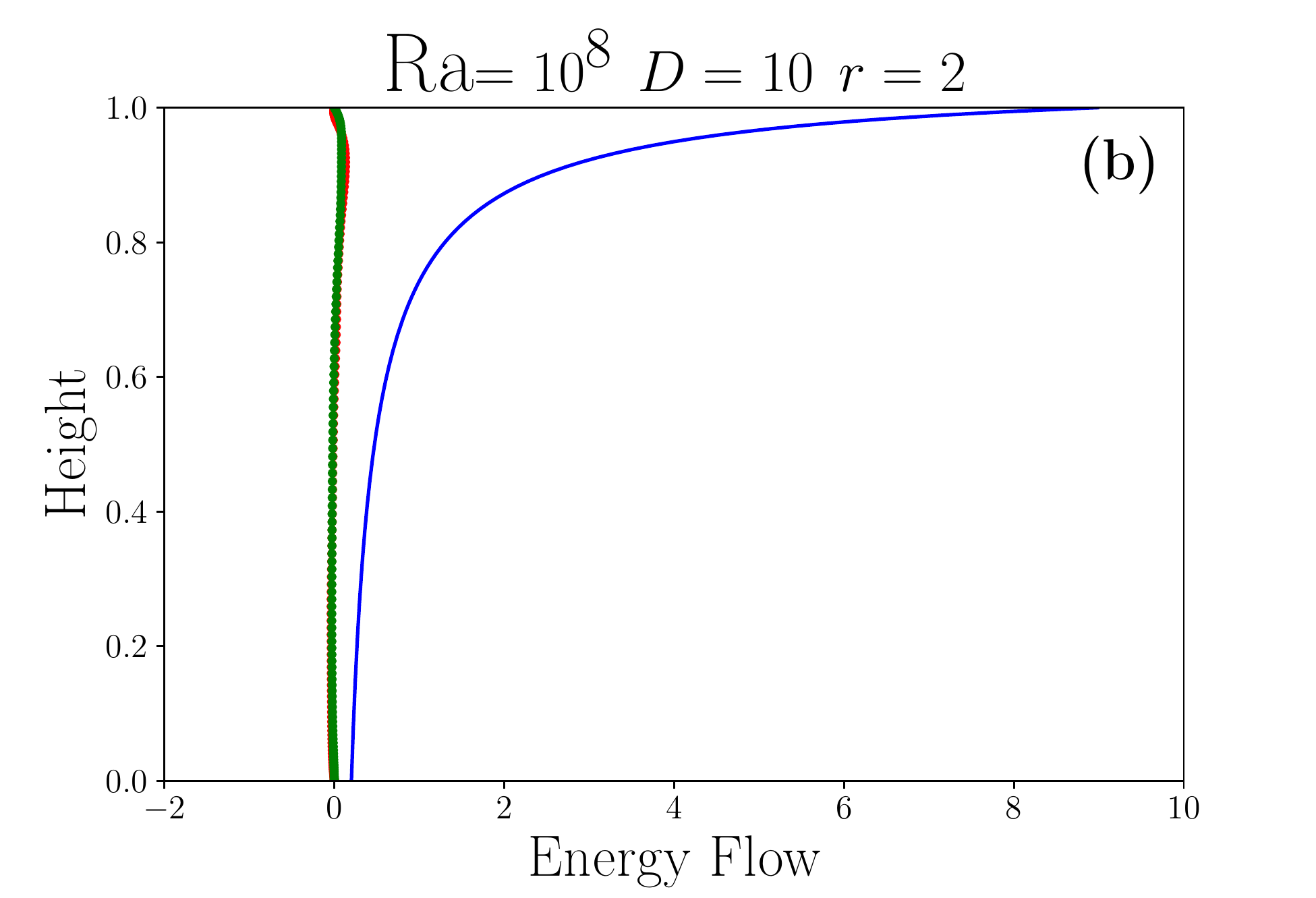}}
\caption{
Energy flow across the convective layer. In panel (a), we plot the total energy flow, averaged on time, in green. The conduction along the non-adiabatic temperature is totally different from the previous cases as now, it carries heat downward near the surface (blue panel (a)). The advection of specific heat (red, panel (a)) dies out at shallow depth. The other minor contributors of energy transport are the conduction along the adiabat (blue, panel (b)), the work flow (green, panel (b)), the contribution due to the fact the enthalpy transport is not restricted to the $C_p T$ term (red, panel (b)). Notice the difference in scale between the two panels.
}
\label{D10r2-3}
\end{figure}

\subsection{Convection with negative superadiabatic Rayleigh number  ($\Di=5$ and $r=1.5$)}
\label{5-1.5}
\begin{figure}
\centerline{\includegraphics[width=18cm,angle=0]{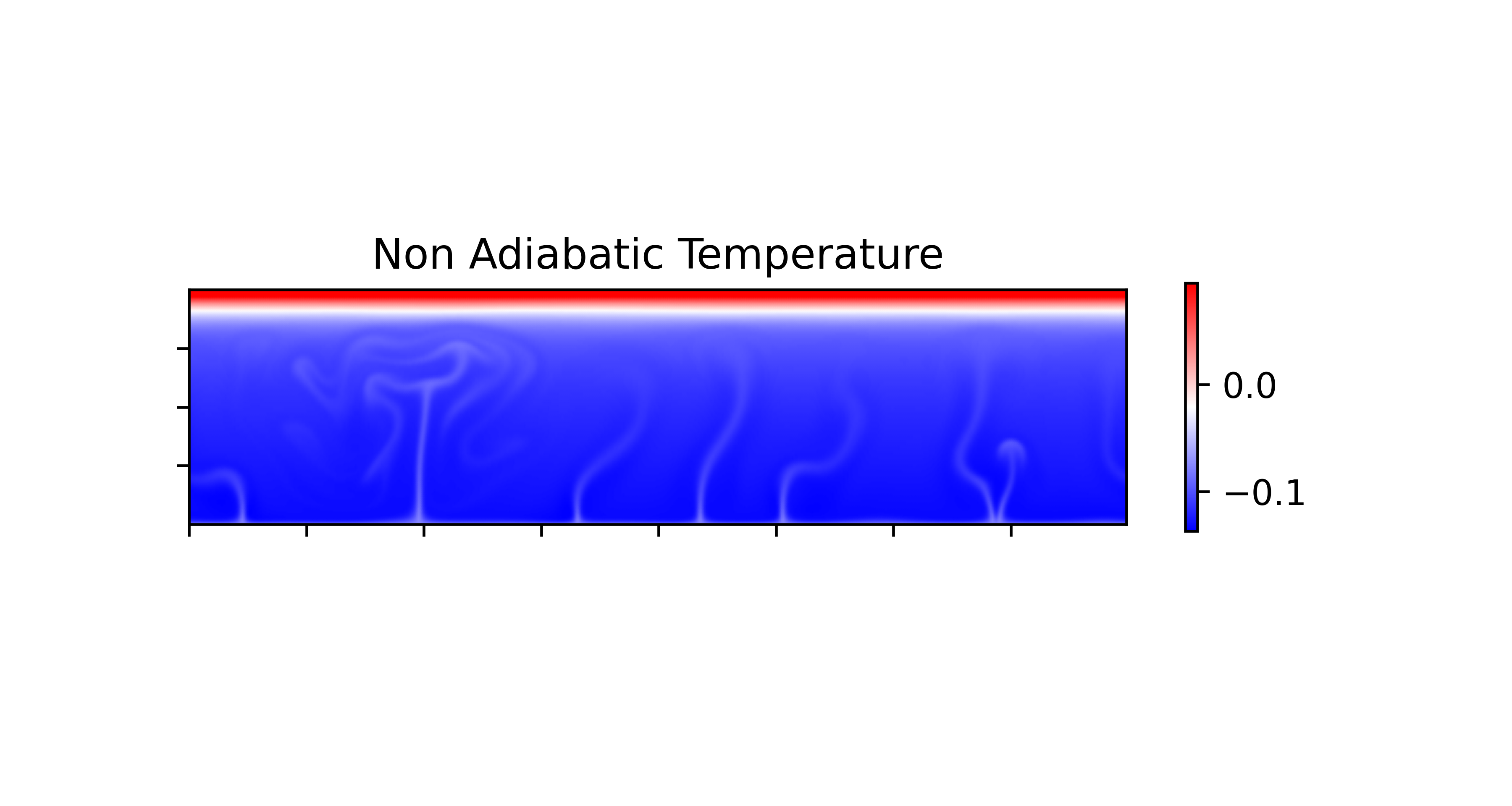}}
\vskip -2cm
\caption{
Snapshot of the superadiabatic temperature in a convection model with a negative Rayleigh number ${\rm Ra}=-10^8$, $\Di=5$ and $r=1.5$. 
As show in Figure \ref{marginal} convection occurs with these parameters.
The color scale is chosen to emphasize the weak rising plumes and saturates near the surface. }
\label{D5r1.5-1}
\end{figure}

\begin{figure}
\centerline{\includegraphics[width=9cm,angle=0]{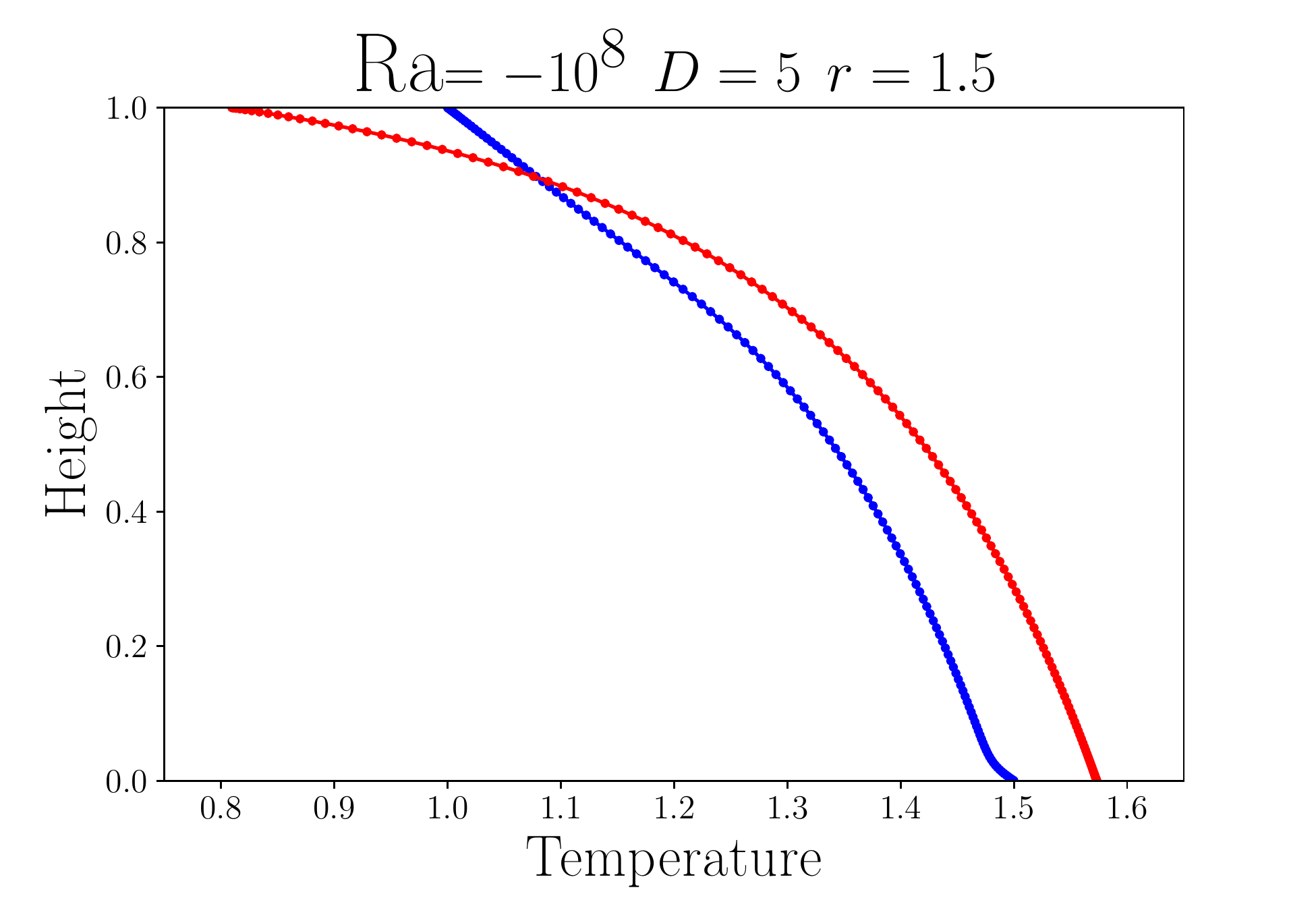}}
\caption{
Temperature profile (blue) and adiabatic (red) in a convection model with ${\rm Ra}=-10^8$, $\Di=5$ and $r=1.5$. The top to bottom temperature ratio is now significantly smaller than the adiabatic temperature difference. Convection occurs with a negative Rayleigh number. A conductive layer in which the temperature gradient is constant fills the upper 20\% of the convective layer but a weak bottom boundary layer allows the emergence of plumes.
}
\label{D5r1.5-2}
\end{figure}

\begin{figure}
\centerline{\includegraphics[width=7.5cm,angle=0]{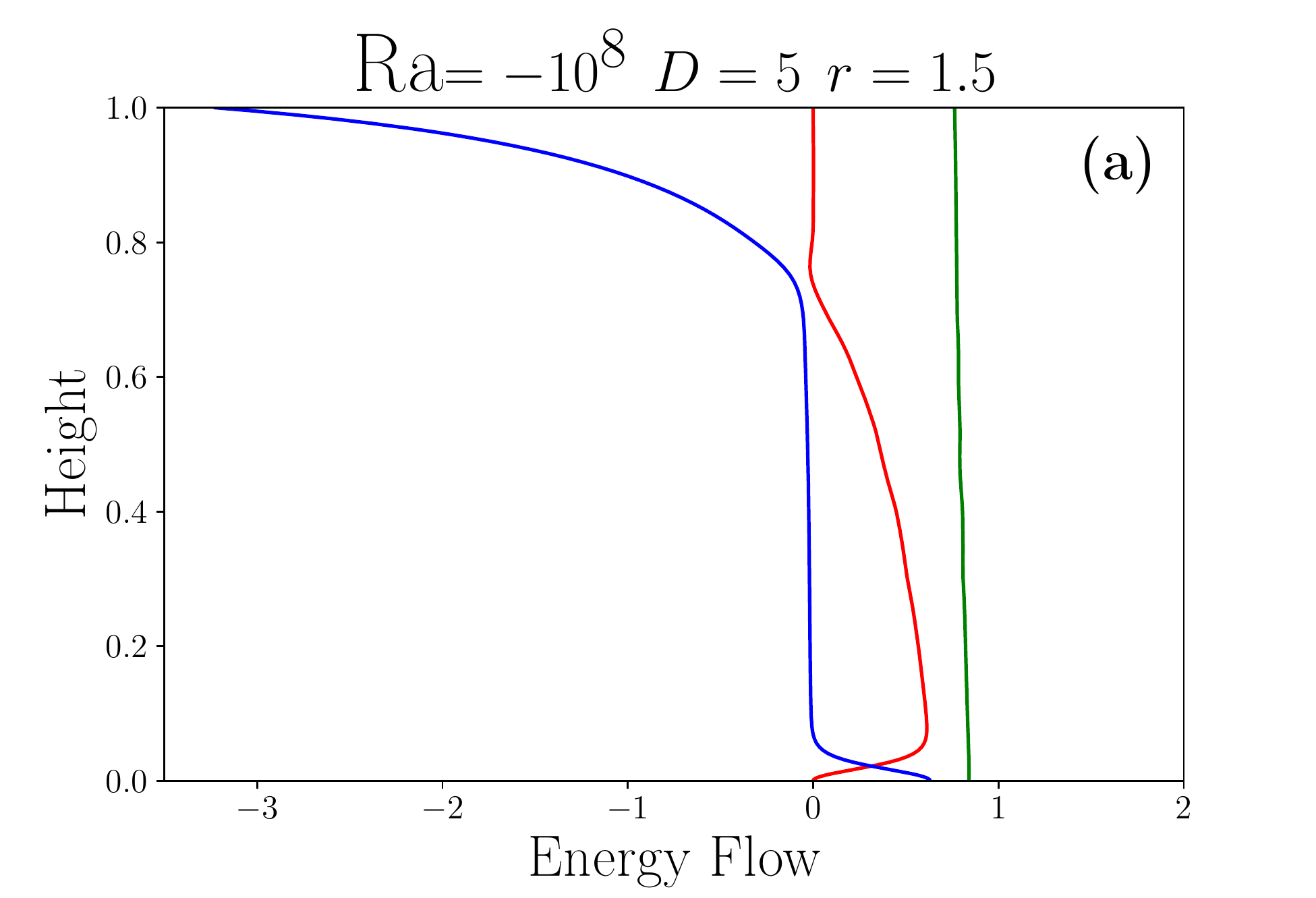}\includegraphics[width=7.5cm,angle=0]{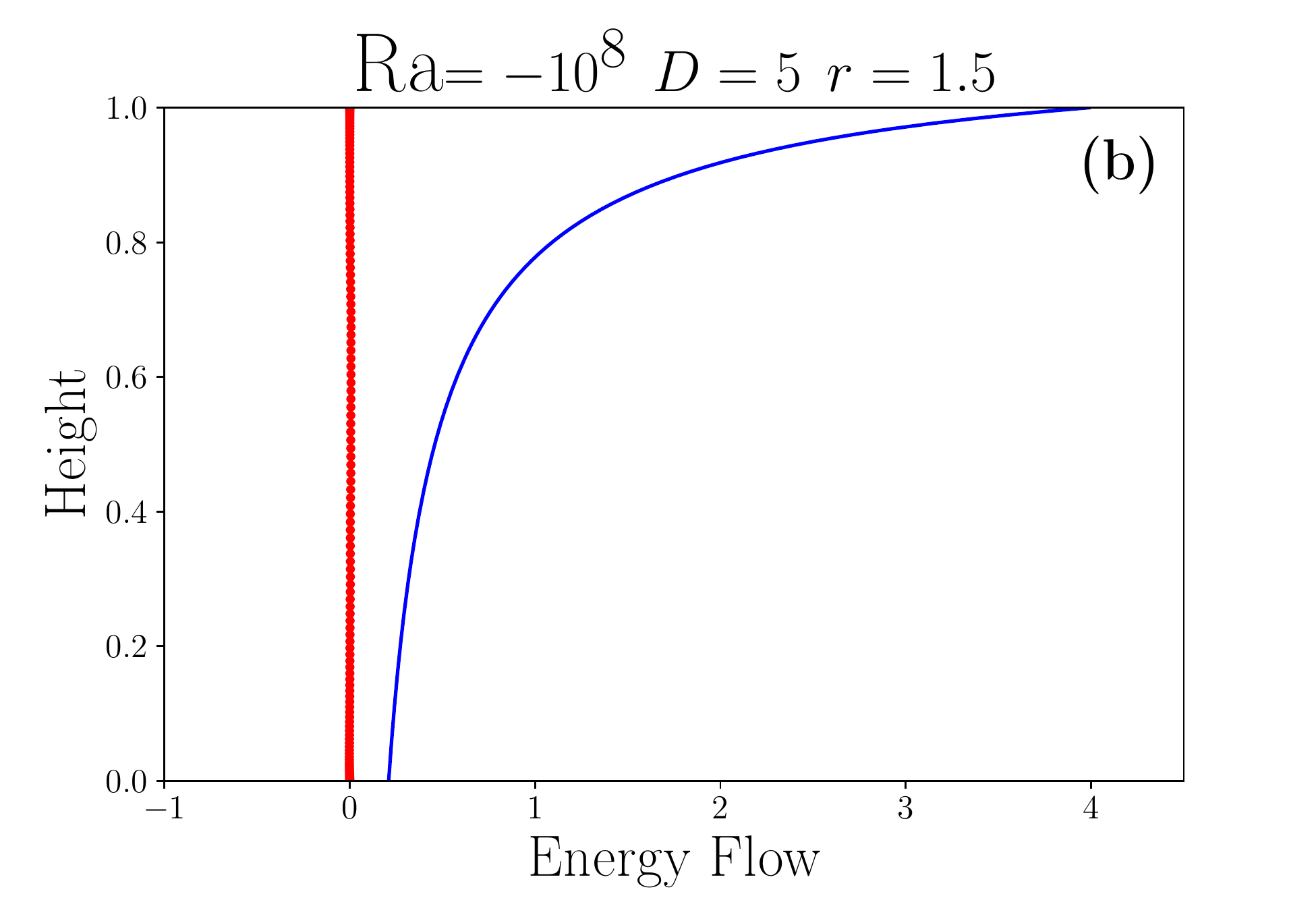}}
\caption{
The conduction terms along the non-adiabatic temperature, in blue panel (a), and along the adiabat, in blue panel (b), largely cancel each other. At depth the energy
is transported by convection (red, panel (a)). The other terms of energy transport are negligible.
}
\label{D5r1.5-3}
\end{figure}

As the marginal stability analysis suggests, convection can also occur with a negative superadiabatic Rayleigh number. This is the case when $\Di=5$ and $r=1.5$ which are parameters that belong to the blue shaded area of the phase diagram of Figure \ref{marginal}. We choose a negative Rayleigh number, ${\rm Ra}=-10^8$, smaller than the negative critical number computed in Figure \ref{marginal}. The resulting convection pattern is typically that shown in Figure \ref{D5r1.5-1}. The shallow layer now appears as a warm (non adiabatic) stable layer beneath which vanish faint rising plumes, warmer than the deep average mantle but colder than the stable surface lid,. The temperature profile of Figure \ref{D5r1.5-2} (blue) is far away from the adiabat (red). The temperature profile is linear in the stable conductive layer ($0.7 \leq z\leq 1$) but a weak hot boundary still remains at the bottom The heat transport components (Figure \ref{D5r1.5-3}) are really dominated by the conductive terms along
the superadiabatic profile (panel (a)) and along the adiabat (panel (b)). These two terms largely cancel each other out but together extract the heat flow transported in the deep layers by convection at the surface (red, panel (a)). The work flow and the difference between enthalpy and specific heat are totally negligible. In this case the Nusselt number is positive with negative numerator and denominator; ${\rm Nu}=12.5$ ($Q=0.85$, $Q_{a}=3.98$, $\Delta T=0.5$, $\Delta T_a=0.76$), but
with a negative Rayleigh number.

\section{Conclusions}

In this paper we have examined how convection is affected by compressibility when the dissipation number which is a measure of the non Boussinesq effects, is large. This situation occurs in Super-Earths, i.e. in solid or liquid planets whose radius much larger than that of the Earth. Such planets are quite commun and the surprising large variety of exoplanets that have been found so far suggests that planets even larger than those considered here (say up to 3 times the radius of the Earth) do exist. The characteristics of compressible flows are controlled by an adiabatic state that provides an approximate reference to the thermodynamic values during developed convection. In a first section we discussed some properties of the adiabatic conditions when the fluid obeys a simple Murnaghan EoS with a Gr\"uneisen parameter decreasing inversely with density. This EoS is quite simple but it faithfully reproduces the behavior of solids and liquids at high pressure and temperature. The consequences of this EoS for the adiabatic density or temperature profiles are quite independent of the nature of the fluid (solid silicate or liquid iron properties can be matched by this generic EoS). This EoS imposes that incompressibility increases rapidly with pressure. The effects of compression are therefore concentrated in the upper layers while the deep layers can hardly be further compressed. This imposes a strong curvature to the adiabatic density and temperature.
The bottom adiabatic temperature is unlikely more than twice the adiabatic temperature at the surface even in large Super-Earths. Although the adiabatic temperature gradient, which plays a major role in convection, may be very large at the surface, it remains moderate at depth and, surprisingly, it decreases, rather than increase, with the dissipation number $\Di$. Another way to think about this is to notice that although $\Di$ estimated from the radius of a planet can be very large at its surface, the Murnaghan EoS requires its mean to be less than the Gr\"uneisen parameter, i.e., typically less than about 1.

We then explore the marginal stability of compressible convection. For our chosen EoS, the strong curvature of the adiabat facilitates convection in the deep layers and promotes conductive heat transport in the shallow layers. Convection can develop in the deep layers even when the total temperature difference between the bottom and top surfaces is equal to or less than the adiabatic temperature difference. Convection with a negative superadiabatic Rayleigh number is therefore possible.

We then explore various cases of developed convection. Our computations are performed with a moderate superadiabatic Rayleigh number ($10^8$ or $-10^8$), various dissipation numbers and various bottom to top temperature ratios. The simulations are performed for a fluid without inertia which is only valid for the creeping convection of planetary mantles. There is of course no indication of which Rayleigh numbers and temperature ratios are
appropriate for Super-Earths. This temperature ratio (and the Rayleigh number) depends on the surface and bottom conditions of the planet's mantle. The former is controlled by the atmosphere composition, the distance to the star, and the internal dynamics of the planet. The latter depends on the mechanisms of formation of the planet, the segregation of the core, the radioactive content of the mantle and the planet's formation age. It is probably unlikely that the imposed temperature ratio across silicated mantles is as small as in the cases of sections \ref{10-2} or \ref{5-1.5} ($\Di\gg r\approx1$) but there are maybe more strange situations than are dreamt in our philosophy. A more commun situation might be provided by the case of section 
\ref{10-10}
 ($\Di\approx r \gg 1$). In that case the top and bottom boundary layers are of very different thicknesses and the concept of a top boundary layer, a lithosphere, may become meaningless as significant conduction along the adiabat can suppress or dampen convection. Heat transport in the case of large dissipation can occur through various combinations of enthalpy transport (which cannot be limited to specific heat), conduction along the adiabatic gradient and along the average superadiabatic temperature gradient and a work flow term. 

In this paper, viscosity has been taken uniform throughout the mantle. The various behaviours that have been described, according to the temperature ratio $r$ and the dissipation number $\Di$, are only due to the response of the EoS to thermal forcing. This means that we do not expect the large scale behaviour of convection to depend on the exact rheology. Of course, the rheology can be different – non newtonian and non uniform – and the precise small-scale flow structures will depend on this, however the overall picture of convection is mainly governed by the EoS.

Although our simulations without inertia are only valid for creeping convection, i.e., for solid state convection, some of our conclusions hold for the case where convection occurs in liquids including
liquid metals. First, as already mentioned, the chosen EoS is probably a much better starting point for correctly expressing the thermodynamical equations that control the flow than is usely done. The characteristics of the adiabatic properties discussed in this paper remain valid for fluids. Of course, other terms, inertia, rotation, electromagnetic effects would have to be possibly added.  For these cases, anelastic approximations should be used for numerical modeling. Starting from a realistic EoS and checking carefully the consistency of the approximations remains necessary.  Situations like in sections \ref{10-2} or \ref{5-1.5} where the dissipation becomes much larger than the temperature ratio are probably common, and might prevail in Earth's core. Note that in these cases, an estimate of a extracted heat flow based on the adiabatic gradient is meaningless because convection does not develop near the surface (see  the difference between the adiabatic transport, blue curve 
of Figure \ref{D5r1.5-3}(b) and the total energy flow, green curve of Figure \ref{D5r1.5-3}(a)). The superadiabatic temperature drives the heat flow down (blue curve of Figure \ref{D5r1.5-3}(a)) and
the surface heat flow remains comparable to what deep convection (red curve of Figure \ref{D5r1.5-3}(a)) is able to carry.

\small

\end{document}